\RequirePackage{fix-cm}
\documentclass[smallcondensed,numbook,envcountsame]{svjour3}     

\usepackage{graphicx}

\usepackage{amsmath,epsfig,amssymb}
\usepackage{cases} 
\usepackage{amssymb,cite}
\usepackage{epsfig}
\usepackage{color}
\usepackage{slashbox}
\usepackage{bm}
\usepackage{cite}
\usepackage{multirow}
\usepackage{slashbox}
\usepackage[utf8]{inputenc}
\usepackage{algorithmic}
\usepackage{algorithm}

\newcommand{\tabincell}[2]{\begin{tabular}{@{}#1@{}}#2\end{tabular}}

\newcommand{\bdm}{\begin{displaymath}}
\newcommand{\edm}{\end{displaymath}}
\newcommand{\be}{\begin{equation}}
\newcommand{\ee}{\end{equation}}
\newcommand{\ba}{\begin{array}}
\newcommand{\ea}{\end{array}}
\newcommand{\bea}{\begin{eqnarray}}
\newcommand{\eea}{\end{eqnarray}}

\def\dfrac#1#2{\frac{\displaystyle #1}{\displaystyle #2}}

\newtheorem{assumption}{Assumption}

\begin{document}

\title{
A New Fully Polynomial Time Approximation Scheme for the Interval Subset Sum Problem\thanks{{Y.-F. Liu was partially supported by NSFC Grants 11671419, 11331012, 11631013, and 11571221. Y.-H. Dai was partially supported by the Key Project of Chinese National Programs for
Fundamental Research and Development Grant 2015CB856000, and NSFC Grants 11631013, 11331012, and 71331001.}}
}

\titlerunning{A New Fully Polynomial Time Approximation Scheme for the ISSP}        

\author{Rui Diao \and Ya-Feng Liu \and Yu-Hong Dai}


\institute{R. Diao, Y.-F. Liu {(Corresponding author)}, and Y.-H. Dai \at State Key Laboratory
of Scientific and Engineering Computing, Institute of Computational
Mathematics and Scientific/Engineering Computing, Academy of
Mathematics and Systems Science, Chinese Academy of Sciences, Beijing 100190, China \\
              \email{diaorui@lsec.cc.ac.cn; yafliu@lsec.cc.ac.cn; dyh@lsec.cc.ac.cn}
}

\date{Received: date / Accepted: date}

\maketitle

\begin{abstract}
The interval subset sum problem (ISSP) is a generalization of the well-known subset sum problem. Given a set of intervals $\left\{[a_{i,1},a_{i,2}]\right\}_{i=1}^n$ and a target integer $T,$ the ISSP is to find a set of integers, at most one from each interval, such that their sum best approximates the target $T$ but {cannot exceed it}. In this paper, we first study the computational complexity of the ISSP. We show that the ISSP is {relatively easy to solve compared to} the 0-1 Knapsack problem (KP). We also identify several subclasses of the ISSP which are polynomial time solvable (with high probability), albeit the problem is generally NP-hard. Then, we propose a new fully polynomial time approximation scheme (FPTAS) for solving the general ISSP problem. The time and space complexities of the proposed scheme are ${\cal O}\left(n \max\left\{1 / \epsilon,\log n\right\}\right)$ and  ${\cal O}\left(n+1/\epsilon\right),$ respectively, {where $\epsilon$ is the relative approximation error}. To the best of our knowledge, the proposed scheme has almost the same time complexity but a significantly lower space complexity compared to the best known scheme. Both the correctness and efficiency of the proposed scheme are validated by numerical simulations. In particular, the proposed scheme successfully solves ISSP instances with $n=100,000$ and $\epsilon=0.1\%$ within {one} second.

\end{abstract}




\section{Introduction}

The subset sum problem (SSP) is a fundamental problem in complexity theory and cryptology. The optimization {formulation} of the SSP is given as follows:

\be \label{prob:ssp}
 \ba{cl}
 \displaystyle \max_{x} & \displaystyle\sum_{i=1}^n a_{i} x_i \\
  \displaystyle \mbox{s.t.} & \displaystyle\sum_{i=1}^n a_{i} x_i \le T, \\
       & x_i \in \{0, 1\},~i=1,2,\ldots,n,\\
 \ea
\ee where $\left\{a_i\right\}_{i=1}^n$ and $T$ are some given positive integers.

The SSP is a famous NP-hard problem \cite{michael1979computers}. Therefore, all exact algorithms for the SSP
are not polynomial unless P=NP. The classical pseudo-polynomial\footnote{{An algorithm that solves a problem is called a {pseudo-polynomial} time algorithm if its time complexity
function is bounded above by a polynomial function related to the numeric value of the input, but exponential in the length of the input.}} algorithm based on the dynamic programming technique for solving the SSP has ${\cal O}(nT)$ time and space complexities. An algorithm with an improved complexity ${\cal O}(n\max\limits_{1\le i\le n}  a_i)$ was proposed by Pisinger \cite{pisinger1999linear}. Various fully polynomial time approximation schemes (FPTAS\footnote{{An algorithm is called an FPTAS for a maximization problem if, for any given instance of the problem and any relative error $\epsilon\in(0,1),$ the algorithm returns a solution value $v^A$ satisfying $v^A\geq \left(1-\epsilon\right)v^*,$ where $v^*$ is the optimal value of the corresponding instance, and its time complexity function is polynomial both in the length of the given data of the problem and in $1/\epsilon.$}}) have also been proposed for the SSP. The first FPTAS for the SSP was proposed by Ibarra and Kim \cite{ibarra1975fast} in 1975, which has a time complexity of being ${\cal O}(n/\epsilon^2)$ and a space complexity of being ${\cal O}\left(n+1/\epsilon^3\right)$.
To the best of our knowledge, the current best FPTAS for the SSP was proposed by Kellerer and Pferschy \cite{kellerer2003efficient}. The time complexity and space complexity of the proposed FPTAS in \cite{kellerer2003efficient} are ${\cal O}(\min\{n/\epsilon, n+1/\epsilon^2\log(1/\epsilon)\})$ and ${\cal O}(n+1/\epsilon),$ respectively. There are also some works focusing on characterizing the easy subclass of the SSP. For instance, in \cite{brickell1984solving,coster1992improved,lagarias1985solving}, both theory and algorithms were proposed for the SSP when $n / \log_2\max\limits_{1\le i \le n} a_i $ is small.

The 0-1 Knapsack problem (KP) is a generalization of the SSP, which has the optimization form as follows:

\be \label{prob:kp}
 \ba{cl}
 \max\limits_{x} & \displaystyle \sum_{i=1}^n v_{i} x_i \\
  \mbox{s.t.} & \displaystyle \sum_{i=1}^n a_{i} x_i \le T, \\
       & x_i \in \{0, 1\},~ i=1,2,\ldots,n,\\
 \ea
\ee {where $\left\{v_i\right\}_{i=1}^n$, $\left\{a_i\right\}_{i=1}^n,$ and $T$ are} some given positive integers.
 When $v_i=a_i$ for all $i=1,2,\ldots,n,$ the 0-1 KP reduces to the SSP.

Various FPTASs have also been proposed for the 0-1 KP. For instance, Lawler \cite{lawler1979fast} proposed {an} FPTAS for the 0-1 KP with time and space complexities being ${\cal O}(n \log(1/\epsilon) + 1/\epsilon^4)$ and ${\cal O}(n+1/\epsilon^3),$ respectively. Later, Magazine and Oguz \cite{magazine1981fully} proposed another PFTAS for the 0-1 KP.  The time and space complexities of the proposed FPTAS in \cite{magazine1981fully} are ${\cal O}(n^2 \log n/\epsilon)$ and ${\cal O}(n /\epsilon),$ respectively. A relatively recent FPTAS, with the time complexity ${\cal O}(n \log n+\min\{n,1/\epsilon \log(1/\epsilon)\}1/\epsilon^2)$ and the space complexity ${\cal O}(n + 1/\epsilon^2),$ was proposed by Kellerer and Pferschy \cite{kellerer1999new}. These FPTASs for the 0-1 KP are summarized in Table \ref{table:kp-issp-ssp-complexity}.




Another generalization of the SSP is the interval subset sum problem (ISSP), which is the focus of this paper.~Mathematically, the ISSP can be formulated as follows:
\be \label{prob:issporigin}
 \ba{cl}
 \max\limits_{x} & \displaystyle \sum_{i=1}^n x_i \\
  \mbox{s.t.} & \displaystyle \sum_{i=1}^n x_i \le T, \\
       & x_i \in \{0\} \cup [a_{i,1}, a_{i,2}],~x_i\in \mathbb{Z},~i=1,2,\ldots, n,\\
 \ea
\ee where $a_{i,2} \ge a_{i,1},~i=1,2,\ldots,n$ are positive integers and $\mathbb{Z}$ denotes the set of integers. When $a_{i,1}=a_{i,2}$ for all $i=1,2,\ldots,n$, the ISSP reduces to the SSP.

The ISSP was first studied by Kothari \emph{et al.} \cite{kothari2005interval} with applications in auction clearing for uniform-price multi-unit auctions. The ISSP has also found wide applications in unit commitment, power generation \cite{carrion2006computationally}, and many others. For instance, in dispatch of the power system, the electric power units need be operated to match the total power load. Each power unit can be chosen to be off or on and the output of each power unit can be adjusted in an interval when it is on. These features can be represented and formulated as the {constraints} in problem \eqref{prob:issporigin}. Currently, the FPTAS proposed by Kothari \emph{et al.} \cite{kothari2005interval} is the only known algorithm designed for solving the ISSP. Both the time complexity and the space complexity of the FPTAS in \cite{kothari2005interval} are ${\cal O}(n/\epsilon)$.

In this paper, we consider the ISSP and propose a new efficient FPTAS for solving it. Compared to the FPTAS for the ISSP in \cite{kothari2005interval}, the proposed FPTAS in this paper has almost the same time complexity but a significantly lower space complexity. Some of the existing FPTASs for the SSP, the ISSP and the 0-1 KP are summarized in Table \ref{table:kp-issp-ssp-complexity}. The {main} contributions of this paper are listed as follows.

\begin{itemize}

\item  [-] The ISSP is shown to be easier than the 0-1 KP in the sense that the ISSP can be equivalently reformulated as a 0-1 KP and therefore any algorithms for the 0-1 KP can be applied to solve the ISSP (see Theorem \ref{thm:issp_and_knapsack});

  \item [-] Some polynomial time solvable subclasses of the ISSP are identified (see Theorems \ref{thm:poly1} and \ref{thm:ktimes-issp}). For instance, it is shown in Theorem \ref{thm:ktimes-issp} that the ISSP is polynomial time solvable when $a_{i,2} \ge 2 a_{i,1}$ for all $i=1,2,\ldots, n$;

     \item [-] By exploiting a new solution structure of the ISSP (see Lemma \ref{lem:neworder}), a new FPTAS is proposed to solve the ISSP. The proposed FPTAS enjoys a significantly lower space complexity compared to the existing one; See Theorems \ref{thm1} and \ref{thm2}.

\item [-] Both the correctness and efficiency of the proposed FPTAS are shown by applying it to solve large scale ISSP instances. In particular, the proposed FPTAS is capable of solving ISSP instances with $n=100,000$ and $\epsilon=0.1\%$ within {one} second; see Section \ref{sec:conclusion}.
\end{itemize}

The rest of this paper is organized as follows. In Section \ref{sec:complexity}, we study the computational complexity of the ISSP.
In Section \ref{sec:algorithm}, we propose a new FPTAS for the ISSP and analyze the time and space complexities of the proposed scheme. Simulation results are presented in Section \ref{sec:experiment} to validate the correctness and efficiency of the proposed FPTAS. The C++ simulation codes are available at [http://bitbucket.org/diaorui/issp]. Finally, some concluding remarks are drawn in Section \ref{sec:conclusion}.

{To streamline the presentation, all proofs of Lemmas/Theorems/Corollaries in this paper are relegated to Appendix A.}

\begin{table}[htbp]
\centering
\caption{Summary of FPTASs for SSP, ISSP, and 0-1 KP}\label{table:kp-issp-ssp-complexity}
\begin{tabular}{|c|c|c|c|}
\hline
Problem & Time Complexity & Space Complexity & Reference \\ \hline
\multirow{1}{*}{SSP}
     & ${\cal O}(\min\{n/\epsilon, n+1/\epsilon^2\log(1/\epsilon)\})$ & ${\cal O}(n+1/\epsilon)$ & \cite{kellerer2003efficient} \\ \hline
\multirow{2}{*}{ISSP} & ${\cal O}(n/\epsilon)$ & ${\cal O}(n/\epsilon)$ & \cite{kothari2005interval} \\ \cline{2-4}
     & ${\cal O}\left(n \max\left\{1 / \epsilon,\log n\right\}\right)$ & ${\cal O}(n + 1/\epsilon)$  & this paper \\ \hline
\multirow{3}{*}{0-1 KP} & ${\cal O}(n \log(1/\epsilon) + 1/\epsilon^4)$ & ${\cal O}(n+1/\epsilon^3)$ & \cite{lawler1979fast} \\ \cline{2-4}
     & ${\cal O}(n^2 \log n/\epsilon)$ & ${\cal O}(n /\epsilon)$ & \cite{magazine1981fully} \\ \cline{2-4}
     & ${\cal O}(n \log n+\min\{n,1/\epsilon \log(1/\epsilon)\}1/\epsilon^2)$ & ${\cal O}(n + 1/\epsilon^2)$ & \cite{kellerer1999new} \\ \hline
\end{tabular}
\end{table}

\section{{Hardness Analysis}}\label{sec:complexity}

The ISSP is generally NP-hard, since it contains the SSP as a special case which is NP-hard. In this section, we first show that the ISSP is easier to solve than the 0-1 KP by proving that the ISSP can be equivalently reformulated as a 0-1 KP. Then, we identify some easy subclasses of the ISSP which can be solved in polynomial time (to global optimality) and therefore clearly delineate the set of computationally tractable problems within the general class of NP-hard ISSPs.

Without loss of generality, we make the following assumption on the inputs of the ISSP throughout this paper.
\begin{assumption}\label{ass:issp}
The inputs of ISSP (\ref{prob:issporigin}) satisfies $T > \displaystyle \max_{1 \le i \le n} a_{i,2}$.
\end{assumption}

If Assumption \ref{ass:issp} is not satisfied, {there must exist} an index $i$ such that $T\leq a_{i,2}.$ In this case, we can either find the solution of the ISSP which is $x_i=T$ and $x_j=0$ for all $j\neq i$ (if $T\geq a_{i,1}$), or remove the corresponding interval $[a_{i,1},a_{i,2}]$ without losing any optimality (if $T< a_{i,1}$).

Next we give an equivalent reformulation of the ISSP.
The variables $\left\{x_i\right\}_{i=1}^n$ in the ISSP are often called semi-continuous\footnote{Strictly speaking, the variables $\left\{x_i\right\}_{i=1}^n$ in the ISSP are not semi-continuous, since $x_i$ in the ISSP can be either zero or integers in the interval $[a_{i,1}, a_{i,2}]$ while the semi-continuous variable $x_i$ can be zero or any continuous value in the corresponding interval. However, as will become clear soon, the intrinsic difficulty of solving the ISSP lies in determining whether $x_i$ should be zero or belong to $[a_{i,1},a_{i,2}].$ Once this is done, it is simple to obtain an optimal solution of the ISSP. Therefore, we actually can drop the constraint $x_i\in\mathbb{Z}$ in the ISSP.} \cite{sun2013recent}. Problems with semi-continuous variables can be equivalently transformed into a mixed integer program by introducing some auxiliary variables. Specifically, we can introduce binary variables $\left\{y_i\right\}_{i=1}^n$ and real variables $\left\{z_i\right\}_{i=1}^n$ satisfying $0 \le z_i \le  (a_{i,2} - a_{i,1})  y_i,~i=1,2,\ldots,n.$ Then, for any $i=1,2,\ldots,n,$ it is simple to verify
$$x_i \in \{0\} \cup [a_{i,1}, a_{i,2}]\Longleftrightarrow x_i =  a_{i,1} y_i  + z_i,~y_i\in\left\{0,1\right\},~0 \le z_i \le  (a_{i,2} - a_{i,1})  y_i.$$ Therefore, ISSP \eqref{prob:issporigin} can be equivalently reformulated as
\be \label{issp}
 \ba{cl}
 \max\limits_{y,z} & \displaystyle \sum_{i=1}^n (a_{i,1} y_i + z_i) \\
  \mbox{\text{s.t.}} & \displaystyle \sum_{i=1}^n (a_{i,1} y_i + z_i) \le T, \\
       & y_i \in \{0, 1\},~i=1,2,\ldots,n,\\
       & 0 \le z_i \le (a_{i,2} - a_{i,1}) y_i,~i=1,2,\ldots,n.\\
 \ea
\ee
We can further eliminate the variables $\left\{z_i\right\}_{i=1}^n$ in the above problem \eqref{issp} and transform it into an equivalent problem with only binary variables.

\begin{theorem}\label{thm:issp_and_knapsack}
ISSP (\ref{issp}) is equivalent to
\be \label{issp_01}
 \ba{cl}
 \max\limits_{y} & \min \left\{ \displaystyle \sum_{i=1}^n a_{i,2} y_i, T \right\} \\
  \mbox{s.t.} &  \displaystyle \sum_{i=1}^n a_{i,1} y_i \le T, \\
       & y_i \in \{ 0, 1 \},~i=1,2,\ldots,n.\\
 \ea
\ee
\end{theorem}

Theorem \ref{thm:issp_and_knapsack} builds a bridge between the ISSP and the 0-1 KP. It shows that any algorithms designed for the 0-1 KP \be \label{kp}
 \ba{cl}
 \max\limits_{y} & \displaystyle \sum_{i=1}^n a_{i,2} y_i  \\
  \mbox{s.t.} &  \displaystyle \sum_{i=1}^n a_{i,1} y_i \le T, \\
       & y_i \in \{ 0, 1 \},~i=1,2,\ldots,n\\
 \ea
\ee can be used to solve the corresponding ISSP and thus the ISSP is easier than the 0-1 KP. In particular, if the optimal value of problem \eqref{kp} is greater than or equal to $T,$  we can obtain a solution of the ISSP from the solution of problem \eqref{kp} (see {Case} {A} of the proof of Theorem \ref{thm:issp_and_knapsack} in Appendix A); otherwise, the two problems share the same solution. Table \ref{table:kp-issp-ssp-complexity} summarizes some known FPTASs for the SSP, the ISSP, and the 0-1 KP, which are consistent with the above analysis, i.e., the difficulty of the ISSP lies between the SSP and the 0-1 KP. Theorem \ref{thm:issp_and_knapsack} also implies that any subclass of the ISSP is polynomial time solvable if the corresponding 0-1 KP problem \eqref{kp} is polynomial time solvable.

Now, we present some polynomial time solvable subclasses of the ISSP.

\begin{theorem}\label{thm:poly1}
 Suppose the inputs of the ISSP satisfy 
 \be \label{condition} T \ge  \left\lceil \dfrac{\max_{1\le i \le n} \{a_{i,1}\} }{\min_{1\le i \le n}\left\{a_{i,2} - a_{i,1}\right\}} \right\rceil  \max\limits_{1\le i \le n} \{ a_{i,1} \}.\ee
Then the ISSP is polynomial time solvable.
\end{theorem}

\begin{theorem}\label{thm:ktimes-issp}
Suppose \be\label{ratio}a_{i,2} \ge c \, a_{i,1},~i=1,2,\ldots, n\ee holds true for some $c>1$ and $T$ obeys the uniform distribution over the interval $\left(\max\limits_{1\le i\le n} a_{i,2}, \displaystyle\sum_{i=1}^n a_{i,2}\right].$ Then, the probability that the ISSP is polynomial time solvable is at least $\min \left\{ 1, 2(1-\dfrac{1}{c}) \right\}.$ Moreover, if \eqref{ratio} holds true with $c\geq2,$ then ISSP (\ref{prob:issporigin}) is polynomial time solvable.
\end{theorem}

Theorems \ref{thm:poly1} and \ref{thm:ktimes-issp} identify two subclasses of the ISSP which are polynomial time solvable. They essentially {indicate} that the ISSP is polynomial time solvable if the length of the intervals are sufficiently large. In particular, Theorem \ref{thm:poly1} shows that the ISSP is polynomial time solvable if $a_{i,2}-a_{i,1}$ is sufficiently large for all $i$ and Theorem \ref{thm:ktimes-issp} shows that the ISSP is polynomial time solvable if $a_{i,2}/a_{i,1}$ is sufficiently large for all $i.$ These results are consistent with our intuition, since there is more freedom to pick an element in a large interval.

\section{A New FPTAS for the ISSP}\label{sec:algorithm}
In this section, we propose a new FPTAS for the ISSP and analyze its time and space complexities. More specifically, we first give a new solution structure of the ISSP in Section \ref{subsec:solution}. By the use of this new solution property, we propose a new FPTAS for the ISSP. Since the proposed FPTAS is rather lengthy and complicated, we first give a high level preview of it in Section \ref{subsec:preview} and then {provide a technical description of it} in Section \ref{subsec:algorithm}. Finally, we analyze the time and space complexities of the proposed FPTAS in Section \ref{subsec:analysis}.

\subsection{A New Solution Structure}\label{subsec:solution}
To present the new structure, we first review some existing results of the solution of the ISSP in \cite{kothari2005interval}.
\begin{definition}[\cite{kothari2005interval}]\label{defn:midrange}
For any feasible solution $\{x_i\}_{i=1}^n$ of ISSP \eqref{prob:issporigin}, if $x_i=a_{i,1},$ then $[a_{i,1}, a_{i,2}]$ is called the left anchored interval; if $x_i=a_{i,2},$ then $[a_{i,1}, a_{i,2}]$ is called the right anchored interval; if $x_i\in(a_{i,1}, a_{i,2}),$ then $[a_{i,1}, a_{i,2}]$ is called the midrange interval.
\end{definition}

\begin{lemma}[\cite{kothari2005interval}]\label{onlyonemidrange}
The ISSP has an optimal solution with at most one midrange interval.
\end{lemma}
\begin{lemma}[\cite{kothari2005interval}]\label{lem:canonicalsolution}
The ISSP has an optimal solution with the following property: if the optimal solution has one midrange interval, then all left anchored intervals precede the midrange interval and all right anchored intervals follow the midrange interval.
\end{lemma}


Now, we present the new solution structure of the ISSP. 
\begin{lemma}\label{lem:neworder}
 Suppose the inputs of the ISSP satisfy $$a_{1,2}-a_{1,1} \le a_{2,2}-a_{2,1} \le \cdots \le a_{n,2}-a_{n,1}.$$
  Then the ISSP has an optimal solution with the following property: if $x_j$ is the midrange element in the solution, then neither the left anchored intervals nor the right anchored intervals follow $[a_{j,1},a_{j,2}].$
\end{lemma}

By exploiting the special structure of the solution of the ISSP in Lemma \ref{lem:canonicalsolution}, an FPTAS was proposed by Kothari \emph{et al.} in \cite{kothari2005interval}. The proposed FPTAS in this paper is based on the new solution structure of the ISSP in Lemma \ref{lem:neworder}. The time and space complexities of the two FPTASs can be found in Table \ref{table:kp-issp-ssp-complexity}.

\subsection{High Level Preview of the Proposed FPTAS}\label{subsec:preview}

{In this subsection, we give a high level preview of the proposed FPTAS. Before doing that, we first present a pseudo-polynomial time dynamic programming algorithm (Algorithm \ref{algo:issp_pseudopoly}) for solving the ISSP. Algorithm \ref{algo:issp_pseudopoly} exploits the special solution structure of the ISSP in Lemma \ref{lem:neworder} and its basic idea is to enumerate the possible midrange interval. More specifically, in Algorithm \ref{algo:issp_pseudopoly}, the set $\Delta_i^*$ in line \ref{line:dynamic} contains all values  $$\left\{\sum_{j\in {\mathcal J}_i} x_j~|~\sum_{j\in {\mathcal J}_i} x_j\leq T,~x_{j}\in\left\{a_{j,1},a_{j,2}\right\},~{\mathcal J}_i\subset \left\{1,2,\ldots,i\right\}\right\}$$ and the only possible midrange interval is $[a_{m,1},a_{m,2}]$ (see line \ref{line:midrange}), where $m$ is the smallest one achieving the maximum value among
$$\left\{\max \{ \delta \in \Delta_{i-1}^* ~|~ \delta \le T - a_{i,1} \}+a_{i,2}\right\}_{i=1}^n.$$
In particular, if $$\max \{ \delta \in \Delta_{m-1}^* ~|~ \delta \le T - a_{m,1} \}+a_{m,2}\geq T,$$ then the target $T$ is achievable. In this case, there is no need to compute $\left\{\Delta_i^*\right\}_{i=m}^n$ (see line \ref{line:goto}).
~Once the only possible midrange interval $[a_{m,1},a_{m,2}]$ is found, the optimal solution $\left\{x_i^*\right\}_{i=1}^n$ of the ISSP can be obtained as follows: the optimal solution $\left\{x_i^*\right\}_{i=1}^{m-1},$ which contributes to generate $\delta^*$ in line \ref{line:deltastar}, can be found by a simple backtracking procedure (see line \ref{line:solution1}); the optimal solution $x_m^*$ is set to be $\min \{  a_{m,2}, T - \delta^* \}$ (see line \ref{line:solution2}); and the optimal solution $\left\{x_i^*\right\}_{i=m+1}^{n}$ are set to be zero (see line \ref{line:solution3}). Therefore, Algorithm \ref{algo:issp_pseudopoly} returns an optimal solution satisfying the property in Lemma \ref{lem:neworder}. {An illustration how Algorithm \ref{algo:issp_pseudopoly} works is given in Appendix B.} Although Algorithm \ref{algo:issp_pseudopoly} can solve the ISSP to global optimality, both of its time complexity (of computing all values in $\left\{\Delta_i^*\right\}_{i=1}^n$) and space complexity (of storing them) are ${\cal O}(nT).$}

\begin{algorithm}\label{algo:issp_pseudopoly} \textbf{Algorithm \ref{algo:issp_pseudopoly}: A Pseudo-Polynomial Time Algorithm for the ISSP}
\begin{algorithmic}[1]
\REQUIRE a set of intervals $\Lambda = \{[a_{i,1}, a_{i,2}]\}_{i=1}^n$ and a target value $T$
\ENSURE the optimal solution $\left\{x_i^*\right\}_{i=1}^n$ and the optimal value $T^*$
\STATE sort the intervals such that $a_{i,2} - a_{i,1} \le a_{i+1,2} - a_{i+1,1},~i=1, 2,\ldots, n-1$
\STATE $T^* \leftarrow 0,$ $\Delta_0^* \leftarrow \emptyset$ {(here $\emptyset$ is the empty set)}
\FOR{$i=1,2,\ldots,n$}
\STATE $\delta^* \leftarrow \max \{ \delta \in \Delta_{i-1}^* ~|~ \delta \le T - a_{i,1} \}$ {\small{(here we define $\max \emptyset = 0$)}} \label{line:forstart}
\IF{$\min\{\delta^* + a_{i,2}, T\} > T^*$}
\STATE $T^* \leftarrow \min\{\delta^* + a_{i,2}, T\}$
\STATE $m \leftarrow  i$\label{line:midrange}
\ENDIF
\IF{$T^*=T$}
\STATE go to line \ref{line:deltastar}\label{line:goto}
\ENDIF
\STATE $\Delta_i^* \leftarrow \left(\{\delta + a_{i,1}, \delta + a_{i,2} ~|~ \delta \in \Delta_{i-1}^* \}  \cup \{ a_{i,1}, a_{i,2} \} \cup  \Delta_{i-1}^* \right) \cap (0,T] $\label{line:dynamic}
\ENDFOR
\STATE $\delta^* \leftarrow \max \{ \delta \in \Delta_{m-1}^* ~|~ \delta \le T - a_{m,1} \}$\label{line:deltastar}
\STATE backtrack to find a subset of $\left\{a_{i,1}, a_{i,2}\right\}_{i=1}^{m-1}$ which contributes to generate $\delta^*$, and then construct the solution $\left\{x_i^*\right\}_{i=1}^{m-1}$ \label{line:solution1}
\STATE $x_m^* \leftarrow \min \{  a_{m,2}, T - \delta^* \}$ \label{line:solution2}
\STATE $x_i^* \leftarrow 0,~i=m+1,\ldots,n$ \label{line:solution3}
\STATE \textbf{return} $T^*$ and $\left\{x_i^*\right\}_{i=1}^n$
\end{algorithmic}
\end{algorithm}


Next, we give a high level preview of the proposed FPTAS (Algorithm \ref{algo:issp-fptas}), which can be obtained by doing some nontrivial modifications to the above Algorithm \ref{algo:issp_pseudopoly}. We remark that the proposed FPTAS for the ISSP can be used to solve the SSP.

One main modification made to Algorithm \ref{algo:issp_pseudopoly} is to partition the interval $(0,T]$ into finitely many subintervals (depending on the given relative error $\epsilon$) and to store only the smallest and largest values lying in the subintervals at each iteration. More specifically, for any given relative error $\epsilon>0,$ Algorithm \ref{algo:issp-fptas} partitions the interval $(0,T]$ into $l=\left\lceil 1/\epsilon\right\rceil$ subintervals $$I_1=(0, \epsilon T],~I_2=(\epsilon T, 2 \epsilon T], \ldots, I_l=(( l - 1)\epsilon T, T]$$
and only stores the smallest value $\delta^-(k)$ and the largest value $\delta^+(k)$ in each subinterval {$I_k$}, $k=1,2,\ldots,l;$ see lines \ref{line:begin_update_delta} -- \ref{line:end_update_delta}.~
This is in sharp contrast to Algorithm \ref{algo:issp_pseudopoly}, where all values in $\left\{\Delta_i^*\right\}_{i=1}^n$ (if the target $T$ is not achievable) or all values in $\left\{\Delta_i^*\right\}_{i=1}^{m-1}$ (if the target $T$ is achievable and $[a_{m,1},a_{m,2}]$ is the only possible midrange interval) are stored. Lemma \ref{lem:relaxeddp-approx} and Corollary \ref{cor:relaxeddp-approx} show that doing so will not {lose} much optimality. If we are only interested in obtaining an approximate objective value but not {in getting} the corresponding solution set, we {can} run Algorithm \ref{algo:issp-fptas} without line \ref{line:call_dc}. Line \ref{line:call_dc} of Algorithm \ref{algo:issp-fptas} aims at recovering a corresponding solution set.

According to lines \ref{line:tildedelta}--\ref{line:endfor} of Algorithm \ref{algo:issp-fptas}, for each $\delta\in\left\{\delta^{-}(k),\delta^{+}(k)\right\}_{k=1}^l,$ there exist $\delta'$ and $a_{i,j}$ such that $\delta=\delta'+a_{i,j},$ where $\delta'$ is either zero or a summation of end points of some subset of $\left\{[a_{k,1},a_{k,2}]\right\}_{k=1}^{i-1}$ and $a_{i,j}$ is the last item contributed to generate $\delta.$ Obviously, to recover the approximate solution, it is useful to store such $i$ and $j.$ These indices are stored in $d_1(\cdot)\in\{1,2,\ldots,n\}$ and $d_2(\cdot)\in\{1,2\}$ in the procedure \textbf{relaxed dynamic programming}. Therefore, one natural way of recovering the approximate solution is to simply backtrack the elements in $\left\{\delta^{-}(k),\delta^{+}(k)\right\}_{k=1}^l$ with the help of $d_1(\cdot)$ and $d_2(\cdot).$ However, there is a potential problem with such a simple backtracking procedure, {since an element in the subinterval $I_k$ which generates an element in the interval $I_j$ with $j>k$ might be updated after considering the item with the index $d_1(j).$} For instance, the above $\delta'$ which contributes to generate $\delta$ might be updated, and thus $\delta'$ does no belong to $\left\{\delta^{-}(k),\delta^{+}(k)\right\}_{k=1}^l$ any more. Therefore, the simple backtracking procedure does not work and it is necessary to make a modification to it in order to reconstruct a correct approximate solution.

{Algorithm \ref{algo:issp-fptas} overcomes the previously mentioned backtracking problem by performing the procedure \textbf{divide and conquer}. Once the procedure \textbf{backtracking} can not continue,~the procedure \textbf{divide and conquer} splits the task of constructing an approximate solution with inputs $\tilde \Lambda$ and $\tilde T$ into two subsets $\tilde \Lambda_1$ and $\tilde \Lambda_2$ of (almost) the same cardinality. The procedure \textbf{relaxed dynamic programming} is then performed for both item sets independently with the target value $\tilde T,$ which returns four reduced achievable arrays $\delta_1^-(\cdot),\,\delta_1^+(\cdot),~\delta_2^-(\cdot),~\delta_2^+(\cdot)$ and the associated interval and end point sets $d_{1,1}(\cdot),~d_{1,2}(\cdot),~d_{2,1}(\cdot),~d_{2,2}(\cdot).$
Lemma \ref{lem:divideinto2parts} guarantees that there exist $u_1 \in \{ 0, \delta_1^-(\cdot), \delta_1^+(\cdot) \}$ and $u_2 \in \{ 0, \delta_2^-(\cdot), \delta_2^+(\cdot) \}$ such that $\tilde T - \epsilon T \le u_1 + u_2 \le \tilde T.$

To find the approximate solution corresponding to values $u_1$ and $u_2$ in the above, the procedure \textbf{backtracking} is first performed for the item set $\tilde \Lambda_1$ with the target value $\tilde T-u_2,$ which reconstructs a part of the solution contributed by $\tilde \Lambda_1$ with value $y_1^B.$ In addition, the procedure \textbf{backtracking} also records the items that have been considered to reconstruct the approximate solution. These items are collected in the set $\Lambda^E$ and will not be considered any more. Obviously, by doing so, no item will appear twice in the approximate solution. Lemma \ref{lem:backtracking} states that this will not lose any optimality in terms of reconstructing an approximate solution. If $y_1^B$ is not {sufficiently} close to $\tilde T-u_2,$ a recursive execution of the procedure \textbf{divide and conquer} for the item set $\tilde \Lambda_1\setminus \Lambda^E$ with the target value $\tilde T-u_2-y_1^B$ finally reconstructs a solution with the value $y_1^B+y_1^{DC},$ which is close enough to $u_1.$ The same can be applied to the item set $\tilde \Lambda_2$ with the target value $\tilde T-y_1^B-y_1^{DC},$ which reconstructs a solution with the value $y_2^B+y_2^{DC}.$  Lemma \ref{dciscorrect} shows that the returned approximate value $y^{DC}=y_1^B+y_1^{DC}+y_2^B+y_2^{DC}$ satisfies $\tilde T-\epsilon T\leq y^{DC}\leq \tilde T.$

Since each execution of the procedure \textbf{divide and conquer} returns at least one item for the solution by backtracking, the depth of its recursion is bounded by {${\cal O}(\log n).$} Therefore, Algorithm \ref{algo:issp-fptas} will terminate after (totally) recursively calling the procedure \textbf{divide and conquer} ${n}$ times. Theorems \ref{thm1} and \ref{thm2} show that Algorithm \ref{algo:issp-fptas} is indeed {an} FPTAS for the ISSP with time and space complexities being ${\cal O}\left(n \max\left\{1 / \epsilon,\log n\right\}\right)$ and ${\cal O}(n+1/\epsilon),$ respectively.

\subsection{Technical Description of the Proposed FPTAS}\label{subsec:algorithm}

In this subsection, we describe the proposed FPTAS for the ISSP in a technical fashion, where we use $I_k$ to denote the $k$-th subinterval, use $\delta^-(k)$ and $\delta^+(k)$ to denote the smallest and largest values in $I_k,$ and use $\Lambda^E$ to denote the set of intervals to be removed during the execution of the algorithm. Algorithm \ref{algo:issp-fptas} below is the proposed FPTAS for the ISSP, which calls the procedure \textbf{divide and conquer}. The procedure \textbf{divide and conquer} further calls the procedures \textbf{relaxed dynamic programming} and \textbf{backtracking} to construct the approximate solution.

\begin{algorithm} \textbf{Algorithm \ref{algo:issp-fptas}: A New FPTAS for the ISSP}
\begin{algorithmic}[1] \label{algo:issp-fptas}
\REQUIRE a set of intervals $\Lambda = \left\{[a_{i,1}, a_{i,2}]\right\}_{i=1}^n$, the target value $T$, and the relative approximation error $\epsilon$
\ENSURE the approximate solution $\left\{x_i^A\right\}_{i=1}^n$ and the approximate value $T^A$
\STATE sort the intervals such that $a_{i,2} - a_{i,1} \le a_{i+1,2} - a_{i+1,1},~i=1,2,\ldots,n-1$ \label{sort}\label{line:begin_update_delta}
\STATE $\hat T \leftarrow 0$,~$\hat \delta \leftarrow 0$
\STATE $\delta^+(i)\leftarrow 0,~\delta^-(i)\leftarrow 0,~i = 1,2,\ldots, l:=\left\lceil\frac{1}{\epsilon}\right\rceil$
\FOR{$i=1,2,\ldots,n$}
\STATE $\bar \delta \leftarrow \max\limits_{1\le k\le l} \{ \delta^-(k),\,\delta^+(k)\,|\,\delta^-(k),\,\delta^+(k) \le T - a_{i,1}  \}$ \label{line:beginfor} 
\IF{$\min\{\bar \delta + a_{i,2}, T\} > \hat T$}
\STATE $\hat T \leftarrow \min\{\bar \delta + a_{i,2}, T\}$
\STATE $\hat \delta \leftarrow \bar \delta$
\STATE $m \leftarrow  i$
\ENDIF
\IF{$T^*=\hat T$}
\STATE go to line \ref{line:lambdaE}
\ENDIF
\STATE $\tilde \Delta \leftarrow \left( \{ \delta^-(k) + a_{i,1} , \delta^+(k) + a_{i,1} ,  \delta^-(k) + a_{i,2} , \delta^+(k) + a_{i,2} \} _{k=1}^{l}  \cup  \{ a_{i,1}, a_{i,2} \} \right)\cap (0, T] $\label{line:tildedelta}
\FOR{each $\tilde \delta \in \tilde \Delta$}
\STATE find $I_k$ that contains $\tilde\delta$
\IF{$\tilde\delta < \delta^-(k)$ or $\delta^-(k) = 0$} \label{line:begin_update_delta}
\STATE $\delta^-(k) \leftarrow \tilde\delta$
\ENDIF
\IF{$\tilde\delta > \delta^+(k)$ or $\delta^+(k) = 0$}
\STATE $\delta^+(k) \leftarrow \tilde\delta$
\ENDIF \label{line:end_update_delta}

\ENDFOR\label{line:endfor}
\ENDFOR
\STATE $\Lambda^E \leftarrow \left\{ [a_{i,1}, a_{i,2}] ~|~ m\le i \le n\right\}$ \label{line:lambdaE}
\STATE $x_i^A \leftarrow 0,~i=1,\ldots,m-1$
\STATE call the procedure \textbf{divide and conquer} $\left(\Lambda \setminus \Lambda^E, \min\left\{\hat \delta + \epsilon T,T-a_{m,1}\right\}\right)$ for the approximate solution $\left\{x_i^A\right\}_{i=1}^{m-1}$ and the approximate value $\hat T^A$ \label{line:call_dc}
\STATE $x_m^A \leftarrow \min \{  a_{m,2}, T - \hat T^A \}$\label{line:m}
\STATE $x_i^A \leftarrow 0,~i=m+1,\ldots,n$\label{line:m+1}
\STATE $T^A \leftarrow \hat T^A + x_m^A$ \label{line:last}
\STATE \textbf{return} $T^A$ and $\left\{x_i^A\right\}_{i=1}^n$
\end{algorithmic}
\end{algorithm}

In {Algorithm \ref{algo:issp-fptas}}, only the largest value $\delta^+(k)$ and the smallest value $\delta^-(k)$ in each subinterval $k=1,2,\ldots,l:=\lceil1/\epsilon\rceil$  are stored; see lines \ref{line:begin_update_delta} -- \ref{line:end_update_delta}. Lemma \ref{lem:relaxeddp-approx} and Corollary \ref{cor:relaxeddp-approx} show that this will not lose much optimality. It will become clear from Theorem \ref{thm1} and the discussions below it why the the last input of the procedure \textbf{divide and conquer} is set to be $\min\left\{\hat \delta + \epsilon T,T-a_{m,1}\right\}$ in line \ref{line:call_dc}. By doing so, Algorithm \ref{algo:issp-fptas} is able to return an (exactly) optimal solution of the ISSP when $\hat \delta+\epsilon T\leq T-a_{m,1}$.

\begin{algorithm*}
\noindent \textbf{Procedure divide and conquer} $(\tilde \Lambda, \tilde T)$
\begin{algorithmic}[1] \label{algo:dc}
\REQUIRE a subset of intervals $\tilde \Lambda \subseteq \{[a_{i,1}, a_{i,2}]\}_{i=1}^n$ and the target value $\tilde T$
\ENSURE the updated approximate solution $\{\hat x_i\}_{i=1}^n$ and the approximate value $y^{DC}$
\STATE split $\tilde \Lambda$ into $\tilde \Lambda_1$ and $\tilde \Lambda_2$, which contain $\left\lceil \dfrac{ | \tilde \Lambda | }{2} \right\rceil$ and $\left\lfloor \dfrac{ | \tilde \Lambda |}{2} \right\rfloor$ elements respectively\label{line:dc1}
\STATE call \textbf{relaxed dynamic programming} $(\tilde \Lambda_1, \tilde T)$ to obtain $\delta_1^-(\cdot), \delta_1^+(\cdot), d_{1,1}(\cdot), d_{1,2}(\cdot)$ \label{line:conquer1}
\STATE call \textbf{relaxed dynamic programming} $(\tilde \Lambda_2, \tilde T)$ to obtain $\delta_2^-(\cdot), \delta_2^+(\cdot), d_{2,1}(\cdot), d_{2,2}(\cdot)$ \label{line:conquer2}
\STATE find $u_1 \in \{ 0, \delta_1^-(\cdot), \delta_1^+(\cdot) \}$ and $u_2 \in \{ 0, \delta_2^-(\cdot), \delta_2^+(\cdot) \}$ such that $\tilde T - \epsilon T \le u_1 + u_2 \le \tilde T$ {\small{(Lemma \ref{lem:divideinto2parts} guarantees the existence of such $u_1$ and $u_2$)}}\label{line:conqueru}

\STATE $y_1^B\leftarrow 0,\,y_1^{DC}\leftarrow 0,\,y_2^B\leftarrow 0,\,y_2^{DC}\leftarrow 0$
\IF{$\tilde T - u_2 > \epsilon T$}
\STATE call \textbf{backtracking} $(\delta_1^-(\cdot), \delta_1^+(\cdot), d_{1,1}(\cdot), d_{1,2}(\cdot), \tilde \Lambda_1, \tilde T - u_2)$ to obtain $y_1^B$ and $\Lambda^E$ \label{line:dc_backtrack1}
\ENDIF
\IF{$\tilde T - u_2 - y_1^B > \epsilon T$}
\STATE call \textbf{divide and conquer} $(\tilde \Lambda_1 \setminus \Lambda^E, \tilde T-u_2-y_1^B)$ to obtain $y_1^{DC}$ \label{line:dc_dc1}
\ENDIF
\IF{$\tilde T-y_1^B-y_1^{DC} > \epsilon T$}\label{line:dc_startif2}
\STATE call \textbf{relaxed dynamic programming} $(\tilde \Lambda_2, \tilde T - y_1^B - y_1^{DC})$ to obtain $\delta_2^-(\cdot),$ $\delta_2^+(\cdot),$ $d_{2,1}(\cdot),$ $d_{2,2}(\cdot)$ \label{line:conquer3}
\STATE call \textbf{backtracking} $(\delta_2^-(\cdot), \delta_2^+(\cdot), d_{2,1}(\cdot), d_{2,2}(\cdot), \tilde \Lambda_2, \tilde T - y_1^B - y_1^{DC})$ to obtain $y_2^B$ and $\Lambda^E$ \label{line:dc_backtrack2}
\ENDIF \label{line:dc_endif2}
\IF{$\tilde T - y_1^B-y_1^{DC}-y_2^B >\epsilon T$}
\STATE call \textbf{divide and conquer} $(\tilde \Lambda_2 \setminus \Lambda^E, \tilde T-y_1^B-y_1^{DC}-y_2^B)$ to obtain $y_2^{DC}$ \label{line:dc_dc2}
\ENDIF
\STATE $y^{DC} \leftarrow y_1^B+y_1^{DC}+y_2^B+y_2^{DC}$
\STATE \textbf{return} $y^{DC}$
\end{algorithmic}
\end{algorithm*}

The procedure \textbf{divide and conquer}, with inputs $(\tilde \Lambda, \tilde T),$ aims at finding an approximate solution $\{\hat x_i\}_{i=1}^n$ and the approximate value $y^{DC}$ from the given subset of intervals $\tilde \Lambda \subseteq \{[a_{i,1}, a_{i,2}]\}_{i=1}^n$ such that $\tilde T-\epsilon T\leq y^{DC}\leq \tilde T.$ Lemma \ref{dciscorrect} establishes that the returned approximate value $y^{DC}=y_1^B+y_1^{DC}+y_2^B+y_2^{DC}$ indeed satisfies $\tilde T-\epsilon T\leq y^{DC}\leq \tilde T.$

\begin{algorithm*}
\noindent\textbf{Procedure relaxed dynamic programming} $(\tilde\Lambda, \tilde T)$
\begin{algorithmic}[1] \label{algo:relaxeddp}
\REQUIRE a subset of intervals $\tilde \Lambda \subseteq \{[a_{i,1}, a_{i,2}]\}_{i=1}^n$ and the target value $\tilde T$
\ENSURE dynamic programming arrays $\delta^-(\cdot), \delta^+(\cdot), d_1(\cdot), d_2(\cdot)$
\STATE $\delta^+(i)\leftarrow 0,~\delta^-(i)\leftarrow 0$, $~i = 1,2,\ldots, \tilde l:=\left\lceil \tilde T/(\epsilon T) \right\rceil$
\FOR{each $i \in \{ i ~|~ [a_{i,1}, a_{i,2}]  \in \tilde \Lambda \}$}
\FOR{$j=1,2$}
\STATE $\tilde \Delta \leftarrow  \left( \{ \delta^-(k) + a_{i,j},\,\delta^+(k) + a_{i,j} \}_{k=1}^{\tilde l}  \cup \{ a_{i,j} \} \right) \cap (0, \tilde T] $
\FOR{each $\tilde \delta \in \tilde \Delta$}
\STATE find $I_k$ that contains $\tilde\delta$
\IF{$\tilde\delta < \delta^-(k)$ or $\delta^-(k) = 0$} \label{line:begin_update_d}
\STATE $\delta^-(k) \leftarrow \tilde\delta,~d_1(\delta^-(k))\leftarrow  i,~d_2(\delta^-(k))\leftarrow j$
\ENDIF
\IF{$\tilde\delta > \delta^+(k)$ or $\delta^+(k) = 0$}
\STATE $\delta^+(k) \leftarrow \tilde\delta,~d_1(\delta^+(k))\leftarrow  i,~d_2(\delta^+(k))\leftarrow j$
\ENDIF \label{line:end_update_d}
\ENDFOR
\ENDFOR
\ENDFOR
\STATE \textbf{return} $\delta^-(\cdot), \delta^+(\cdot), d_1(\cdot), d_2(\cdot)$
\end{algorithmic}
\end{algorithm*}

The procedure \textbf{relaxed dynamic programming}, with inputs $(\tilde \Lambda, \tilde T),$ is a subroutine used in the procedure \textbf{divide and conquer} to recover the approximate solution. In the procedure \textbf{relaxed dynamic programming}, besides the largest value $\delta^+(k)$ and the smallest value $\delta^-(k)$ in each subinterval $k=1,2,\ldots,\tilde l:=\left\lceil \tilde T/(\epsilon T) \right\rceil,$ the index and the end point of the last interval which generates $\delta\in\left\{\delta^+(k),\delta^-(k)\right\}_{k=1}^{\tilde l}$ are also stored in $d_1(\cdot)\in\left\{1,2,\ldots,\tilde l\right\}$ and $d_2(\cdot)\in\left\{1,2\right\}.$ These information will be used in the procedure \textbf{backtracking} to recover the approximate solution.

\begin{algorithm*}
\noindent \textbf{Procedure backtracking} $(\delta^-(\cdot), \delta^+(\cdot), d_1(\cdot), d_2(\cdot), \tilde \Lambda, \tilde T)$
\begin{algorithmic}[1] \label{algo:backtracking}
\REQUIRE dynamic programming arrays $\delta^-(\cdot), \delta^+(\cdot), d_1(\cdot), d_2(\cdot),$ a subset of intervals $\tilde \Lambda \subseteq \{[a_{i,1}, a_{i,2}]\}_{i=1}^n$, and the target value $\tilde T$
\ENSURE the updated approximate solution $\{\hat x_i\}_{i=1}^n,$ the set of intervals $\Lambda^E,$ and the partial approximate value $y$
\STATE $u\leftarrow \max\{\delta^+(j), \delta^-(j) ~|~  \delta^+(j)\le \tilde T, \delta^-(j) \le \tilde T \}$ \label{line:init_u}
\STATE $y\leftarrow 0$
\REPEAT
\STATE $i\leftarrow d_1(u),\,j\leftarrow d_2(u)$ \label{line:begin_main_backtrack}
\STATE $\hat x_i \leftarrow a_{i,j}$
\STATE $\Lambda^E \leftarrow \Lambda^E \cup \left\{ [a_{k,1}, a_{k,2}] \in \tilde \Lambda ~|~  k \ge i \right\} $
\STATE $y \leftarrow y + a_{i,j}$
\STATE $u \leftarrow u - a_{i,j}$ \label{line:end_main_backtrack}
\IF{$u > 0$} \label{line:begin_speedup_backtrack}
\STATE find $I_k$ that contains $u$
\IF{$\delta^+(k) + y \le \tilde T$ and $d_1(\delta^+(k)) < i$}
\STATE $u \leftarrow \delta^+(k)$
\ELSIF {$\delta^-(k)+y \ge \tilde T - \epsilon T $ and $d_1(\delta^-(k)) < i$}
\STATE $u \leftarrow \delta^-(k)$
\ELSE
\STATE $u \leftarrow 0$
\ENDIF
\ENDIF \label{line:end_speedup_backtrack}
\UNTIL {$u=0$}
\STATE \textbf{return} $y$ and $\Lambda^E$
\end{algorithmic}
\end{algorithm*}

Taking the outputs of the procedure \textbf{relaxed dynamic programming} $\delta^-(\cdot)$, $\delta^+(\cdot)$, $d_1(\cdot),$ and $d_2(\cdot)$ as inputs, the procedure \textbf{backtracking} backtracks the intervals which contribute to generate the largest value in $\{\delta^+(j), \delta^-(j) ~|~  \delta^+(j)\le \tilde T, \delta^-(j) \le \tilde T \}$ (see line \ref{line:init_u}) with the help of $d_1(\cdot)$ and $d_2(\cdot)$ and reconstructs a partial approximate solution. The procedure \textbf{backtracking} stops if for some $u\in I_k,$ the solution value conditions $\delta^+(k) + y \le \tilde T$ and $\delta^-(k)+y \ge \tilde T - \epsilon T$ do not hold true, or $\delta^+(k) + y \le \tilde T$ holds true but $u$ is updated by some later interval with index $d_1(\delta^+(k)) \geq i;$ or $\delta^-(k)+y \ge \tilde T - \epsilon T$ holds true but $u$ is updated by some later interval with index $d_1(\delta^-(k)) \geq i.$ Actually, lines \ref{line:begin_speedup_backtrack} -- \ref{line:end_speedup_backtrack} are to speed up the procedure \textbf{backtracking}. It will become clear from Lemma \ref{lem:backtracking} and the discussions below it that Algorithm \ref{algo:issp-fptas} can still recover an approximate solution of the ISSP if there are no lines \ref{line:begin_speedup_backtrack} --
\ref{line:end_speedup_backtrack} in the procedure \textbf{backtracking}.

{An illustration how Algorithm \ref{algo:issp-fptas} works is given in Appendix B.}

\subsection{Worst-Case Time and Space Complexity Analysis}\label{subsec:analysis}

For simplicity of our analysis, we introduce the following notation. Let $\Delta^*_{\tilde \Lambda}$ denote the full dynamic programming arrays computed from the intervals in $\tilde\Lambda,$ and $\Delta_{\tilde \Lambda}$ the relaxed dynamic programming arrays computed from the intervals in $\tilde\Lambda$. In particular, when $\tilde \Lambda=\left\{1,2,\ldots,i\right\},$ we denote them by $\Delta^*_i$ and $\Delta_i,$ respectively. Throughout this section, we shall also use the term ``$\Delta^*_{\tilde \Lambda}$ associated with a target $\tilde T$'' to denote the set $\left\{\delta\,|\,\delta\in\Delta^*_{\tilde \Lambda},\,\delta\leq \tilde T\right\}.$ The same applies to $\Delta_{\tilde \Lambda},$ $\Delta^*_i,$ and $\Delta_i.$

With the above notation, it is obvious to see that $\left\{\delta\in\Delta_{n}^*\,|\,\delta\leq T\right\}$ is the optimal value of the ISSP. According to lines \ref{line:tildedelta} -- \ref{line:end_update_delta} of Algorithm \ref{algo:issp-fptas},
we know $\Delta_i\subseteq \Delta^*_i$ for all $i=1,2,\ldots,n$ and $\Delta_{\tilde \Lambda}\subseteq \Delta^*_{\tilde \Lambda}$ for all $\tilde \Lambda\subseteq \left\{1,2,\ldots,n\right\}.$

The next lemma (Lemma \ref{lem:relaxeddp-approx}) plays a fundamental role in analyzing the proposed FPTAS for the ISSP. It says that each element computed by the full dynamic programming procedure can be well approximated (with a bounded error) by some element computed by the relaxed dynamic programming procedure. More specifically, Lemma \ref{lem:relaxeddp-approx} shows that, for any $\delta\in \Delta_{\tilde\Lambda}^*,$ there exists an $\underline\delta\in\Delta_{\tilde\Lambda}$ such that $0\leq\delta-\underline\delta\leq \epsilon T.$

\begin{lemma}\label{lem:relaxeddp-approx}
For each $\delta \in \Delta_i^*$ associated with the target $\tilde T$, one of the following two statements is true:
\begin{enumerate}
\item There exist $\underline\delta,\,\overline\delta \in \Delta_i$ such that $\underline\delta \le \delta \le \overline\delta$ and $\overline\delta - \underline\delta \le \epsilon T.$
\item There exists $\underline\delta \in \Delta_i$ such that $\tilde T - \epsilon T \le \underline\delta \le \delta \le \tilde T$.
\end{enumerate}
\end{lemma}

By setting $\delta$ to be the optimal value of the ISSP in Lemma \ref{lem:relaxeddp-approx}, we have the following corollary.
\begin{corollary}\label{cor:relaxeddp-approx}
Suppose $\delta^*$ is the optimal value of the ISSP with inputs $\tilde \Lambda$ and $\tilde T.$ Then, there exists $\delta \in \Delta_{\tilde \Lambda}$ such that either $\delta = \delta^*$ or $\tilde T - \epsilon T \le  \delta \le \delta^* \le \tilde T $ holds true.
\end{corollary}


The following lemma (Lemma \ref{lem:divideinto2parts}) shows the existence of $u_1$ and $u_2$ in line \ref{line:conqueru} of the procedure \textbf{divide and conquer}.
\begin{lemma}\label{lem:divideinto2parts}
 Suppose there exists $\delta \in \Delta_{\tilde \Lambda}$ such that $\tilde T - \epsilon T \le \delta \le \tilde T$ when the procedure \textbf{divide and conquer} starts. Then, after the splitting of $\tilde \Lambda$ into $\tilde \Lambda_1$ and $\tilde \Lambda_2$, there must exist $u_1 \in \{0\} \cup \Delta_{\tilde \Lambda_1}$ and $u_2 \in \{0\} \cup \Delta_{\tilde \Lambda_2}$ such that $\tilde T - \epsilon T \le u_1 + u_2 \le \tilde T$.
\end{lemma}

Lemma \ref{lem:divideinto2parts} establishes the existence of $u_1 \in \{0\} \cup \Delta_{\tilde \Lambda_1}$ and $u_2 \in \{0\} \cup \Delta_{\tilde \Lambda_2}$ such that $\tilde T - \epsilon T \le u_1 + u_2 \le \tilde T.$ In fact, such $u_1$ and $u_2$ might not be unique; see the remark in Appendix B.  In our implementation of the procedure \textbf{divide and conquer}, we use the following strategy, with both time and space complexities being ${\cal O}(1/\epsilon),$ to find a pair of $u_1$ and $u_2$ satisfying $\tilde T - \epsilon T \le u_1 + u_2 \le \tilde T.$ Without loss of generality, suppose $\Delta_{\tilde \Lambda_1}=\left\{\delta_1^{1},\delta_1^2,\ldots,\delta_1^{2l}\right\}$ and $\Delta_{\tilde \Lambda_2}=\left\{\delta_2^{1},\delta_2^2,\ldots,\delta_2^{2l}\right\},$ where $\delta_1^{i}\leq \delta_1^{i+1}$ and $\delta_2^{i}\leq \delta_2^{i+1}$ for all $i=1,2,\ldots,2l-1$ and $l\leq \left\lceil1/\epsilon\right\rceil.$ For convenience, we also define $\delta_1^{0}=\delta_2^{0}=0.$ Let $i=0$ and $j=2l,$ $u_1=\delta_1^0,$ and $u_2=\delta_2^{2l},$ then repeatedly do the following until $u_1+u_2\in[\tilde T-\epsilon T, \tilde T]:$ if $u_1+u_2< \tilde T-\epsilon T,$ let $u_1=\delta_1^{i+1}$ and increment $i$ by $1;$ if $u_1+u_2> \tilde T,$ let $u_2=\delta_2^{j-1}$ and decrement $j$ by $1.$ It is simple to verify that the above strategy returns a pair of desired $u_1$ and $u_2$ and its time and space complexities are ${\cal O}(1/\epsilon).$

The following lemma (Lemma \ref{lem:backtracking}) states that the procedure \textbf{backtracking} can successfully backtrack a part of the approximate solution by using the relaxed dynamic programming arrays. This is in sharp contrast to the pseudo-polynomial time algorithm (Algorithm \ref{algo:issp_pseudopoly}) for the ISSP, where the solution is backtracked by using the full dynamic programming arrays. We remark that the space complexity of storing the relaxed dynamic programming arrays is ${\cal O}(1/\epsilon)$ and the space complexity of storing the full dynamic programming arrays is ${\cal O}(nT).$

\begin{lemma}\label{lem:backtracking}
Suppose there exists $\delta' \in  \Delta^*_{\tilde \Lambda}$ such that $\tilde T - \epsilon T \le  \delta'  \le \tilde T$ when the procedure \textbf{backtracking} starts and $y$ is its output. Then, there exists $\delta \in \{ 0 \} \cup \Delta^*_{\tilde \Lambda \setminus \Lambda^E}$ such that $\tilde T - \epsilon T$ $ \le y + \delta $ $\le \tilde T.$ 
\end{lemma}

{It is worthwhile remarking that lines \ref{line:begin_speedup_backtrack} --
\ref{line:end_speedup_backtrack} are for speeding up the procedure \textbf{backtracking}. More specifically, the procedure \textbf{backtracking} equipped with  lines \ref{line:begin_speedup_backtrack} --
\ref{line:end_speedup_backtrack} could potentially backtrack more than one step. In contrast, the procedure \textbf{backtracking}, without lines \ref{line:begin_speedup_backtrack} --
\ref{line:end_speedup_backtrack}, can only backtrack one step. Lines \ref{line:begin_speedup_backtrack} --
\ref{line:end_speedup_backtrack} will not affect the time and space complexities of the proposed FPTAS, which will become more clear in the following Theorem \ref{thm2}.

\begin{lemma}\label{dciscorrect}
 Suppose there exists $\delta \in \Delta^*_{\tilde \Lambda}$ such that $\tilde T - \epsilon T \le \delta \le \tilde T$ when the procedure \textbf{divide and conquer} starts. Then, the output $y^{DC}$ of the procedure \textbf{divide and conquer} satisfies $\tilde T - \epsilon T \le y^{DC} \le \tilde T.$
\end{lemma}

We are now ready to present the main results of this section.
\begin{theorem}\label{thm1}
Algorithm \ref{algo:issp-fptas} returns an $(1-\epsilon)$-approximate solution of the ISSP.
\end{theorem}

Note that in line \ref{line:call_dc} of Algorithm \ref{algo:issp-fptas}, the last input of the procedure \textbf{divide and conquer} is set to be $\min\left\{\hat \delta + \epsilon T,T-a_{m,1}\right\}.$ This trick makes Algorithm \ref{algo:issp-fptas} return an (exactly) optimal solution of the ISSP when $\hat \delta+\epsilon T\leq T-a_{m,1};$ see Case A of the proof of Theorem \ref{thm1} in Appendix A. Actually, the result presented in Theorem \ref{thm1} still holds true if $\min\left\{\hat \delta + \epsilon T,T-a_{m,1}\right\}$ is replaced with $\hat \delta=\min\left\{\hat \delta,T-a_{m,1}\right\},$ where the last equality is due to $\hat \delta\leq T-a_{m,1}.$ However, if so, Algorithm \ref{algo:issp-fptas} would not enjoy the nice property of returning the (exactly) optimal solution when $\hat \delta+\epsilon T\leq T-a_{m,1}.$

\begin{theorem}\label{thm2}
The time complexity of Algorithm \ref{algo:issp-fptas} is ${\cal O}\left(n \max\left\{1 / \epsilon,\log n\right\}\right)$, and the space complexity is ${\cal O}(n + 1/\epsilon)$.
\end{theorem}

We compare the proposed FPTAS and the one in \cite{kothari2005interval} in terms of the time and space complexities; see Table \ref{table:kp-issp-ssp-complexity}. The time complexity of the two FPTASs are comparable to each other. This is because $\log n$ is generally smaller than $1/\epsilon.$ However, the space complexity of the proposed FPTAS is significantly lower than the one in \cite{kothari2005interval}. The significant improvement in the space complexity makes the proposed FPTAS possible to solve large scale ISSP instances on a limited memory machine, which is generally impossible for the FPTAS in \cite{kothari2005interval}.

\section{Numerical Experiments}\label{sec:experiment}

We implemented the proposed FPTAS for solving the ISSP with C++.  Our numerical experiments were done on a personal computer with Ubuntu 10.04.2 operating system, Intel  Core i7 CPU,  8 GB memory, and the source code is compiled with GCC 4.4.3.

Since there is no available test set for the ISSP, we tested the proposed FPTAS on the test set for the SSP, with some modifications.

\noindent {Instance} A: $a_{i,1} = a_{i,2} = 2^{k+n+1}+2^{k+i}+1$ and $T=\left\lfloor \dfrac{1}{2}\sum\limits_{i=1}^n a_{i,2} \right\rfloor$, where $k=\lfloor \log_2(n)\rfloor;$

\noindent {Instance} B: $a_{i,1} = a_{i,2} = n(n+1)+i$ and $T=\left\lfloor \dfrac{n-1}{2}  \right\rfloor n (n+1) + \dfrac{n (n-1)}{2};$

\noindent {Instance} C: $a_{i,2}$ is an integer generated by uniformly randomly sampling in $[1,10^{14}]$, $a_{i,1}=\left\lfloor\dfrac{a_{i,2}}{c}\right\rfloor,$ and $T=3\times 10^{14}$, where $c \ge 1$ is a given parameter;

\noindent {Instance} D: $a_{i,2}$ is an integer generated by uniformly randomly sampling in $[1,10^{14}]$, $a_{i,1}=\left\lfloor\dfrac{a_{i,2}}{c_i}\right\rfloor,$ and $T=3\times 10^{14},$ where $c_i$ is a real number generated by uniformly randomly sampling in $[1, C]$ with a given parameter $C \ge 1.$

Instances A and B \cite{chvatal1980hard} are the SSPs. They are used to test whether the proposed FPTAS (Algorithm \ref{algo:issp-fptas}) for the ISSP can correctly and efficiently solve the degenerating problems. The optimal solutions of these two instances are easy to obtain but very hard to compute by the branch and bound method \cite{chvatal1980hard}.
Instances C and D are randomly generated. They are used to test the correctness and efficiency of the proposed FPTAS for solving the ISSP.

Table  \ref{issp-probA} summarizes the numerical results of applying the proposed FPTAS to solve Instance A. The parameter $n$ in Instance A can not be very large; otherwise $a_{i,1}, a_{i,2},$ and $T$ will be extremely large. From Table  \ref{issp-probA}, we can observe that the returned relative errors of all tested SSP instances are strictly less than the preselected parameter $\epsilon$. In particular, when $\epsilon=0.1\%$, the returned relative errors of all tested SSP instances are zero, which implies that the proposed FPTAS successively solves all tested SSP instances to global optimality. These numerical results show the correctness of the proposed FPTAS for solving the SSP.

\begin{table}[htbp]
\centering
\caption{Numerical Results of Instance A}\label{issp-probA}
\begin{tabular}{|c|c|c|c|c|c|c|}
\hline
      & \multicolumn{2}{c|}{$\epsilon=10\%$}  & \multicolumn{2}{c|}{$\epsilon=1\%$}  & \multicolumn{2}{c|}{$\epsilon=0.1\%$} \\ \hline
 $n$  & relative error & time(s) & relative error & time(s) & relative error & time(s) \\ \hline
 $10$   &  $0.000\%$  &  $<0.001$   & $0.000\%$ & $<0.001$  & $0.000\%$  & $<0.001$ \\ \hline
$15$   &  $1.515\%$  &  $<0.001$   & $0.342\%$ & $<0.001$  & $0.000\%$  & $0.001$ \\ \hline
$20$   &  $0.000\%$  &  $<0.001$   & $0.000\%$ & $<0.001$  & $0.000\%$  & $0.002$ \\ \hline
$25$   &  $9.616\%$  &  $<0.001$   & $0.059\%$ & $<0.001$  & $0.000\%$  & $0.003$ \\ \hline
$30$   &  $9.690\%$  &  $<0.001$   & $0.000\%$ & $<0.001$  & $0.000\%$  & $0.004$ \\ \hline
$35$   &  $5.558\%$  &  $<0.001$   & $0.347\%$ & $<0.001$  & $0.000\%$  & $0.005$ \\ \hline

\end{tabular}
\end{table}

Table \ref{issp-probB} summarizes the numerical results of applying the proposed FPTAS to solve Instance B. Again, the returned relative errors of all tested SSP instances are not greater than the given tolerance $\epsilon,$
which means that the proposed FPTAS can solve SSP correctly as claimed in Theorem \ref{thm1}.
From Table \ref{issp-probB}, we can observe that the computational time of the proposed FPTAS grows (roughly) linearly with $n$ when $\epsilon$ is fixed and also (roughly) linearly with $1/\epsilon$ when $n$ is fixed. This matches with the time complexity ${\cal O}\left(n\max\left\{1/\epsilon,\log n\right\}\right)$ of the proposed FPTAS, since $1/\epsilon$ and $\log n$ are comparable to each other for the tested case $\epsilon=10\%$ and $1/\epsilon\gg\log n$ for the tested cases $\epsilon=1\%$ and $\epsilon=0.1\%.$~
In particular, it takes the proposed FPTAS less than $0.8$ seconds to solve the SSP instance with $n=10,000$ and $\epsilon=0.1\%.$ From the above numerical results, we can conclude that the proposed FPTAS can efficiently solve the SSP. 

\begin{table}[htbp]
\centering
\caption{Numerical Results of Instance B}\label{issp-probB}
\begin{tabular}{|c|c|c|c|c|c|c|}
\hline
      & \multicolumn{2}{c|}{$\epsilon=10\%$}  & \multicolumn{2}{c|}{$\epsilon=1\%$}  & \multicolumn{2}{c|}{$\epsilon=0.1\%$} \\ \hline
 $n$  & relative error & time(s) &  relative error & time(s) & relative error & time(s)  \\ \hline
 $10$   &  $0.844\%$  &  $<0.001$   & $0.000\%$ & $<0.001$  & $0.000\%$  & $<0.001$ \\ \hline
 $50$   &  $8.353\%$  &  $<0.001$   & $0.071\%$ & $0.001$  & $0.000\%$  & $0.002$ \\ \hline
$100$   &  $8.166\%$  &  $<0.001$   & $0.019\%$ & $0.002$  & $0.000\%$  & $0.004$ \\ \hline
$500$   &  $9.637\%$  &  $0.002$   & $0.902\%$ & $0.008$  & $0.000\%$  & $0.015$ \\ \hline
$1,000$   &  $9.818\%$  &  $0.001$   & $0.851\%$ & $0.008$  & $0.000\%$  & $0.049$ \\ \hline
$5,000$   &  $9.964\%$  &  $0.007$   & $0.970\%$ & $0.042$  & $0.080\%$  & $0.400$ \\ \hline
$10,000$   &  $9.982\%$  &  $0.014$   & $0.985\%$ & $0.085$  & $0.080\%$  & $0.795$ \\ \hline
\end{tabular}
\end{table}

Table \ref{issp-probC} summarizes the numerical results of applying the proposed PFTAS to solve Instance C. The results in Table \ref{issp-probC} are obtained by averaging over $100$ randomly generated ISSP instances for each fixed $n$ and $c$. The worst results among these $100$ tested instances ({in terms of the relative error and the computational time, respectively}) are also reported in parentheses in Table \ref{issp-probC}. Since the optimal value of Instance C is difficult to obtain, we set it to be the target $T$ when calculating the relative errors. The relative errors computed in this way are obviously larger than or equal to the ``true'' relative errors. It can be observed from Table \ref{issp-probC} that the worst relative errors of all tested instances (and thus the average relative errors) are not greater than the desired relative error $\epsilon,$  which shows the correctness of the proposed FPTAS for solving the ISSP. Table \ref{issp-probC} also demonstrates the efficiency of the proposed FPTAS when applied to solve large scale instances of Instance C. The proposed FPTAS is capable of returning an approximate solution of the tested ISSP instances with $n=100,000$ and $\epsilon=0.1\%$ within $0.1$ second in average.
\begin{table}[htbp]
\centering
\caption{Numerical Results of Instance C}\label{issp-probC}
{\small{
\begin{tabular}{|c|c|c|c|c|c|c|c|}
\hline
    &  & \multicolumn{2}{c|}{$\epsilon=10\%$}  & \multicolumn{2}{c|}{$\epsilon=1\%$}  & \multicolumn{2}{c|}{$\epsilon=0.1\%$} \\ \hline
 $n$ & $c$  & relative error & time(s) & relative error & time(s) & relative error & time(s) \\ \hline
 & $1.5$  & \tabincell{c}{$7.158\%$\\$(9.218\%)$}  & \tabincell{c}{$0.000$\\$(0.001)$}  & \tabincell{c}{$0.076\%$\\$(0.209\%)$}  & \tabincell{c}{$0.001$\\$(0.001)$}  & \tabincell{c}{$0.000\%$\\$(0.000\%)$}  & \tabincell{c}{$0.006$\\$(0.006)$} \\ \cline{2-8}
$1,000$ & $1.3$  & \tabincell{c}{$7.874\%$\\$(9.415\%)$}  & \tabincell{c}{$0.000$\\$(0.001)$}  & \tabincell{c}{$0.308\%$\\$(0.470\%)$}  & \tabincell{c}{$0.001$\\$(0.001)$}  & \tabincell{c}{$0.000\%$\\$(0.000\%)$}  & \tabincell{c}{$0.006$\\$(0.008)$} \\ \cline{2-8}
 & $1.1$  & \tabincell{c}{$8.029\%$\\$(9.771\%)$}  & \tabincell{c}{$0.000$\\$(0.001)$}  & \tabincell{c}{$0.643\%$\\$(0.783\%)$}  & \tabincell{c}{$0.001$\\$(0.001)$}  & \tabincell{c}{$0.000\%$\\$(0.000\%)$}  & \tabincell{c}{$0.005$\\$(0.007)$} \\ \hline
 & $1.5$  & \tabincell{c}{$7.498\%$\\$(9.614\%)$}  & \tabincell{c}{$0.002$\\$(0.003)$}  & \tabincell{c}{$0.581\%$\\$(0.648\%)$}  & \tabincell{c}{$0.002$\\$(0.002)$}  & \tabincell{c}{$0.000\%$\\$(0.000\%)$}  & \tabincell{c}{$0.014$\\$(0.018)$} \\ \cline{2-8}
$5,000$ & $1.3$  & \tabincell{c}{$8.187\%$\\$(9.733\%)$}  & \tabincell{c}{$0.002$\\$(0.002)$}  & \tabincell{c}{$0.696\%$\\$(0.750\%)$}  & \tabincell{c}{$0.002$\\$(0.002)$}  & \tabincell{c}{$0.000\%$\\$(0.000\%)$}  & \tabincell{c}{$0.014$\\$(0.016)$} \\ \cline{2-8}
 & $1.1$  & \tabincell{c}{$8.033\%$\\$(9.893\%)$}  & \tabincell{c}{$0.002$\\$(0.003)$}  & \tabincell{c}{$0.848\%$\\$(0.901\%)$}  & \tabincell{c}{$0.002$\\$(0.004)$}  & \tabincell{c}{$0.000\%$\\$(0.000\%)$}  & \tabincell{c}{$0.014$\\$(0.015)$} \\ \hline
 & $1.5$  & \tabincell{c}{$7.373\%$\\$(9.732\%)$}  & \tabincell{c}{$0.004$\\$(0.005)$}  & \tabincell{c}{$0.665\%$\\$(0.746\%)$}  & \tabincell{c}{$0.003$\\$(0.004)$}  & \tabincell{c}{$0.000\%$\\$(0.000\%)$}  & \tabincell{c}{$0.020$\\$(0.021)$} \\ \cline{2-8}
$10,000$ & $1.3$  & \tabincell{c}{$8.144\%$\\$(9.831\%)$}  & \tabincell{c}{$0.004$\\$(0.005)$}  & \tabincell{c}{$0.755\%$\\$(0.820\%)$}  & \tabincell{c}{$0.003$\\$(0.004)$}  & \tabincell{c}{$0.000\%$\\$(0.000\%)$}  & \tabincell{c}{$0.020$\\$(0.022)$} \\ \cline{2-8}
 & $1.1$  & \tabincell{c}{$8.168\%$\\$(9.921\%)$}  & \tabincell{c}{$0.004$\\$(0.005)$}  & \tabincell{c}{$0.892\%$\\$(0.927\%)$}  & \tabincell{c}{$0.004$\\$(0.005)$}  & \tabincell{c}{$0.009\%$\\$(0.029\%)$}  & \tabincell{c}{$0.020$\\$(0.021)$} \\ \hline
 & $1.5$  & \tabincell{c}{$7.304\%$\\$(9.880\%)$}  & \tabincell{c}{$0.020$\\$(0.027)$}  & \tabincell{c}{$0.701\%$\\$(0.883\%)$}  & \tabincell{c}{$0.017$\\$(0.024)$}  & \tabincell{c}{$0.000\%$\\$(0.000\%)$}  & \tabincell{c}{$0.049$\\$(0.053)$} \\ \cline{2-8}
$50,000$ & $1.3$  & \tabincell{c}{$7.954\%$\\$(9.917\%)$}  & \tabincell{c}{$0.021$\\$(0.027)$}  & \tabincell{c}{$0.764\%$\\$(0.918\%)$}  & \tabincell{c}{$0.018$\\$(0.026)$}  & \tabincell{c}{$0.008\%$\\$(0.020\%)$}  & \tabincell{c}{$0.049$\\$(0.051)$} \\ \cline{2-8}
 & $1.1$  & \tabincell{c}{$7.667\%$\\$(9.965\%)$}  & \tabincell{c}{$0.023$\\$(0.026)$}  & \tabincell{c}{$0.900\%$\\$(0.967\%)$}  & \tabincell{c}{$0.019$\\$(0.025)$}  & \tabincell{c}{$0.053\%$\\$(0.067\%)$}  & \tabincell{c}{$0.053$\\$(0.062)$} \\ \hline
 & $1.5$  & \tabincell{c}{$7.570\%$\\$(9.911\%)$}  & \tabincell{c}{$0.037$\\$(0.054)$}  & \tabincell{c}{$0.749\%$\\$(0.916\%)$}  & \tabincell{c}{$0.033$\\$(0.049)$}  & \tabincell{c}{$0.010\%$\\$(0.019\%)$}  & \tabincell{c}{$0.073$\\$(0.077)$} \\ \cline{2-8}
$100,000$ & $1.3$  & \tabincell{c}{$7.999\%$\\$(9.941\%)$}  & \tabincell{c}{$0.042$\\$(0.050)$}  & \tabincell{c}{$0.795\%$\\$(0.942\%)$}  & \tabincell{c}{$0.038$\\$(0.050)$}  & \tabincell{c}{$0.035\%$\\$(0.043\%)$}  & \tabincell{c}{$0.074$\\$(0.086)$} \\ \cline{2-8}
 & $1.1$  & \tabincell{c}{$7.416\%$\\$(9.974\%)$}  & \tabincell{c}{$0.045$\\$(0.050)$}  & \tabincell{c}{$0.914\%$\\$(0.977\%)$}  & \tabincell{c}{$0.041$\\$(0.055)$}  & \tabincell{c}{$0.066\%$\\$(0.077\%)$}  & \tabincell{c}{$0.084$\\$(0.101)$} \\ \hline
\end{tabular}
}}
\end{table}

\begin{table}[htbp]
\centering
\caption{Numerical Results of Instance D}\label{issp-probD}
{\small{
\begin{tabular}{|c|c|c|c|c|c|c|c|}
\hline
    &  & \multicolumn{2}{c|}{$\epsilon=10\%$}  & \multicolumn{2}{c|}{$\epsilon=1\%$}  & \multicolumn{2}{c|}{$\epsilon=0.1\%$} \\ \hline
 $n$ & $C$  & relative error & time(s) & relative error & time(s) & relative error & time(s) \\ \hline
 & $1.5$  & \tabincell{c}{$5.353\%$\\$(9.881\%)$}  & \tabincell{c}{$0.000$\\$(0.001)$}  & \tabincell{c}{$0.319\%$\\$(0.945\%)$}  & \tabincell{c}{$0.000$\\$(0.001)$}  & \tabincell{c}{$0.003\%$\\$(0.049\%)$}  & \tabincell{c}{$0.001$\\$(0.003)$} \\ \cline{2-8}
$1,000$ & $1.3$  & \tabincell{c}{$4.830\%$\\$(9.866\%)$}  & \tabincell{c}{$0.000$\\$(0.000)$}  & \tabincell{c}{$0.247\%$\\$(0.926\%)$}  & \tabincell{c}{$0.000$\\$(0.001)$}  & \tabincell{c}{$0.004\%$\\$(0.057\%)$}  & \tabincell{c}{$0.001$\\$(0.003)$} \\ \cline{2-8}
 & $1.1$  & \tabincell{c}{$4.891\%$\\$(9.851\%)$}  & \tabincell{c}{$0.000$\\$(0.001)$}  & \tabincell{c}{$0.239\%$\\$(0.931\%)$}  & \tabincell{c}{$0.001$\\$(0.002)$}  & \tabincell{c}{$0.006\%$\\$(0.086\%)$}  & \tabincell{c}{$0.002$\\$(0.005)$} \\ \hline
 & $1.5$  & \tabincell{c}{$5.986\%$\\$(9.979\%)$}  & \tabincell{c}{$0.001$\\$(0.001)$}  & \tabincell{c}{$0.392\%$\\$(0.974\%)$}  & \tabincell{c}{$0.001$\\$(0.003)$}  & \tabincell{c}{$0.017\%$\\$(0.082\%)$}  & \tabincell{c}{$0.002$\\$(0.005)$} \\ \cline{2-8}
$5,000$ & $1.3$  & \tabincell{c}{$4.576\%$\\$(9.805\%)$}  & \tabincell{c}{$0.001$\\$(0.001)$}  & \tabincell{c}{$0.362\%$\\$(0.991\%)$}  & \tabincell{c}{$0.001$\\$(0.003)$}  & \tabincell{c}{$0.018\%$\\$(0.087\%)$}  & \tabincell{c}{$0.002$\\$(0.007)$} \\ \cline{2-8}
 & $1.1$  & \tabincell{c}{$5.232\%$\\$(9.949\%)$}  & \tabincell{c}{$0.001$\\$(0.002)$}  & \tabincell{c}{$0.314\%$\\$(0.986\%)$}  & \tabincell{c}{$0.002$\\$(0.004)$}  & \tabincell{c}{$0.017\%$\\$(0.091\%)$}  & \tabincell{c}{$0.004$\\$(0.010)$} \\ \hline
 & $1.5$  & \tabincell{c}{$6.375\%$\\$(9.943\%)$}  & \tabincell{c}{$0.001$\\$(0.002)$}  & \tabincell{c}{$0.324\%$\\$(0.985\%)$}  & \tabincell{c}{$0.002$\\$(0.005)$}  & \tabincell{c}{$0.023\%$\\$(0.091\%)$}  & \tabincell{c}{$0.003$\\$(0.009)$} \\ \cline{2-8}
$10,000$ & $1.3$  & \tabincell{c}{$6.511\%$\\$(9.988\%)$}  & \tabincell{c}{$0.001$\\$(0.002)$}  & \tabincell{c}{$0.397\%$\\$(0.979\%)$}  & \tabincell{c}{$0.002$\\$(0.004)$}  & \tabincell{c}{$0.019\%$\\$(0.088\%)$}  & \tabincell{c}{$0.003$\\$(0.011)$} \\ \cline{2-8}
 & $1.1$  & \tabincell{c}{$5.191\%$\\$(9.967\%)$}  & \tabincell{c}{$0.002$\\$(0.004)$}  & \tabincell{c}{$0.324\%$\\$(0.982\%)$}  & \tabincell{c}{$0.003$\\$(0.007)$}  & \tabincell{c}{$0.017\%$\\$(0.096\%)$}  & \tabincell{c}{$0.007$\\$(0.014)$} \\ \hline
 & $1.5$  & \tabincell{c}{$6.637\%$\\$(9.969\%)$}  & \tabincell{c}{$0.005$\\$(0.007)$}  & \tabincell{c}{$0.415\%$\\$(0.987\%)$}  & \tabincell{c}{$0.006$\\$(0.010)$}  & \tabincell{c}{$0.033\%$\\$(0.098\%)$}  & \tabincell{c}{$0.010$\\$(0.021)$} \\ \cline{2-8}
$50,000$ & $1.3$  & \tabincell{c}{$6.148\%$\\$(9.982\%)$}  & \tabincell{c}{$0.005$\\$(0.008)$}  & \tabincell{c}{$0.372\%$\\$(0.991\%)$}  & \tabincell{c}{$0.007$\\$(0.014)$}  & \tabincell{c}{$0.032\%$\\$(0.099\%)$}  & \tabincell{c}{$0.012$\\$(0.034)$} \\ \cline{2-8}
 & $1.1$  & \tabincell{c}{$5.695\%$\\$(9.991\%)$}  & \tabincell{c}{$0.006$\\$(0.009)$}  & \tabincell{c}{$0.376\%$\\$(0.995\%)$}  & \tabincell{c}{$0.009$\\$(0.019)$}  & \tabincell{c}{$0.026\%$\\$(0.095\%)$}  & \tabincell{c}{$0.021$\\$(0.054)$} \\ \hline
 & $1.5$  & \tabincell{c}{$6.477\%$\\$(9.994\%)$}  & \tabincell{c}{$0.010$\\$(0.013)$}  & \tabincell{c}{$0.389\%$\\$(0.990\%)$}  & \tabincell{c}{$0.012$\\$(0.023)$}  & \tabincell{c}{$0.030\%$\\$(0.097\%)$}  & \tabincell{c}{$0.018$\\$(0.046)$} \\ \cline{2-8}
$100,000$ & $1.3$  & \tabincell{c}{$6.256\%$\\$(9.993\%)$}  & \tabincell{c}{$0.011$\\$(0.014)$}  & \tabincell{c}{$0.438\%$\\$(0.999\%)$}  & \tabincell{c}{$0.013$\\$(0.022)$}  & \tabincell{c}{$0.029\%$\\$(0.098\%)$}  & \tabincell{c}{$0.022$\\$(0.055)$} \\ \cline{2-8}
 & $1.1$  & \tabincell{c}{$6.245\%$\\$(9.983\%)$}  & \tabincell{c}{$0.012$\\$(0.017)$}  & \tabincell{c}{$0.431\%$\\$(0.980\%)$}  & \tabincell{c}{$0.017$\\$(0.029)$}  & \tabincell{c}{$0.025\%$\\$(0.098\%)$}  & \tabincell{c}{$0.035$\\$(0.096)$} \\ \hline

\end{tabular}
}}
\end{table}

It is worthwhile remarking that the choice of the parameter $c$ in Instance C actually affects the efficiency of the proposed FPTAS. As we can observe from Table \ref{issp-probC}, as the parameter $c$ in Instance C decreases, the average computational time of using the proposed FPTAS to solve the corresponding ISSP instances slightly increases. This becomes obvious for the ISSP instances with $n\geq 50,000.$ The above observation is consistent with Theorem \ref{thm:ktimes-issp}, which basically says that the ISSP becomes more difficult to solve as the parameter $c$ there decreases.

Table \ref{issp-probD} summarizes the numerical results of applying the proposed FPTAS to solve Instance D.
Instance D is similar to Instance C but more general. Therefore, it can better evaluate the performance of the proposed FPTAS.
The same observations as on Instance C can be made on Instance D. Therefore, we can conclude that the proposed FPTAS can solve the ISSP correctly and efficiently.

\section{Concluding Remarks}\label{sec:conclusion}
In this paper, we considered the NP-hard ISSP, which is a generalization of the well-known SSP. We first showed that the ISSP can be equivalently reformulated as a 0-1 KP. This reformulation implies that the ISSP is easier to solve than the 0-1 KP, since any algorithms designed for the 0-1 KP can be used to solve the ISSP. Moreover, we identified several polynomial time solvable subclasses of the ISSP and thus clearly delineated a set of computationally tractable problems within the general class of NP-hard ISSPs. Then, by exploiting a new solution structure of the ISSP, we proposed a new FPTAS for it. Compared to the currently best known FPTAS, the proposed one has a comparable time complexity but a significantly lower space complexity. Numerical results demonstrate the correctness and efficiency of the proposed FPTAS.

The cardinality constrained ISSP \cite{kothari2005interval} is an extension of the ISSP with an extra cardinality constraint $\|x\|_0\leq k_{\max},$ where $\|x\|_0$ denotes the number of nonzero elements in $x.$ The proposed FPTAS for the ISSP in this paper can be modified to solve the cardinality constrained ISSP. The major modification is that $k_{\max}$ of dynamic programming arrays $\tilde \Delta_k,~k=1,2,\ldots,k_{\max}$ need to be introduced in the proposed FPTAS for the ISSP, where $\tilde \Delta_k$ is the same as the dynamic programming array in Algorithm \ref{algo:issp-fptas} except that each element in $\tilde \Delta_k$ is a summation of exactly $k$ end points of the intervals $\left\{[a_{i,1},a_{i,2}]\right\}_{i=1}^n.$ In this way, the above modified algorithm is able to efficiently deal with the cardinality constraint. By using the same argument as in Theorems \ref{thm1} and \ref{thm2}, it can be shown that the above modified algorithm is an FPTAS for the cardinality constrained ISSP and its time and space complexities is ${{\cal O}\left(n \max\left\{k_{\max}/\epsilon, \log n\right\}\right)}$ and ${\cal O}\left(n + k_{\max}/\epsilon\right),$ respectively.

\section*{Appendix A: Proofs of Lemmas/Theorems/Corollaries}
\subsection*{Proof of Theorem \ref{thm:issp_and_knapsack}}
\begin{proof}
We prove the theorem by dividing the proof into two cases, i.e., whether there exists binary $\left\{\bar y_i\right\}_{i=1}^n$ such that \be\label{lowerupper}\sum_{i=1}^n a_{i,1} \bar y_i \le T \le \sum\limits_{i=1}^n a_{i,2} \bar y_i.\ee

{{ Case A}:} there exists binary $\left\{\bar y_i\right\}_{i=1}^n$ such that \eqref{lowerupper} holds true. Without loss of generality, assume \be\label{barI} \bar y_i=1,~i=1,2,\ldots,\bar I;~\bar y_i=0,~i=\bar I+1,\ldots, n.\ee In this case, we claim that both the optimal value of problems \eqref{issp} and \eqref{issp_01} are equal to $T.$ Let us argue the above claim holds. First, it follows from \eqref{lowerupper} that $\left\{\bar y_i\right\}_{i=1}^n$ is feasible to problem \eqref{issp_01} and the optimal value of problem \eqref{issp_01} is equal to $T.$ Now, we evaluate the objective function of problem \eqref{issp} at point $\left\{\bar y_i\right\}_{i=1}^n:$ $$g(z):=\sum_{i=1}^n \left(a_{i,1}\bar y_i+z_i\right)=\sum_{i=1}^{\bar I}a_{i,1}+\sum_{i=1}^{\bar I}z_i,$$ where the last equality is due to the assumption \eqref{barI}. Since $z_i$ can take any value in the interval $[0, a_{i,2}-a_{i,1}]$ for each $i=1,2,\dots,\bar I,$ we know that $g(z)$ can take any value between ${\left [\sum_{i=1}^{\bar I} a_{i, 1}, \, \sum_{i=1}^{\bar I} a_{i,2}\right]}.$ Combining this, \eqref{lowerupper}, and the constraint $\sum_{i=1}^n (a_{i,1} y_i + z_i) \le T,$ we know that the optimal value of problem \eqref{issp} is {exactly} equal to $T.$ {As a matter of fact}, an integer solution $\left\{\bar z_i\right\}_{i=1}^n$ to achieve the optimal value $T$ can be found as follows. {Calculate}
 $$v^i=\sum_{l=1}^{i} a_{l,2}+\sum_{l=i+1}^{\bar I} a_{l,1},~i=0,1,\ldots,\bar I-1.$$ Then there must exist an index $i^*\in\{0,1,\ldots,\bar I-1\}$ such that $v^{i^*}<T\leq v^{i^*+1},$ and $\left\{\bar z_i\right\}_{i=1}^n$ to achieve the optimal value $T$ is given by
  \begin{equation}\label{zi}\bar z_i=\left\{
\begin{array}{cl} & \displaystyle a_{i,2}-a_{i,1},~i=1,\ldots,i^*,\\
     & T-v^{i^*},~i=i^*+1,\\
    &  0,~i=i^*+2,\ldots,n.
    \end{array}\right.\end{equation}

We now show that there is a correspondence between the solutions of problems \eqref{issp} and \eqref{issp_01}. On one hand, for any solution $\left\{y_i^*,z_i^*\right\}_{i=1}^n$ of problem \eqref{issp} achieving the optimal value $T$ (from the above claim), we have $\sum_{i=1}^n a_{i,1} y^*_i \le \sum_{i=1}^n (a_{i,1} y^*_i + z^*_i) = T$, and $\sum\limits_{i=1}^n a_{i,2} y^*_i \ge \sum\limits_{i=1}^n (a_{i,1} y^*_i + z^*_i) = T.$ This immediately shows that $\left\{y_i^*\right\}_{i=1}^n$ is a solution of problem \eqref{issp_01}. On the other hand, suppose that $\left\{y_i^*\right\}_{i=1}^n$ is a solution of problem \eqref{issp_01} achieving the optimal value $T$ (from the above claim). Then, there must hold $\sum\limits_{i=1}^n a_{i,1} y^*_i \le T \le \sum\limits_{i=1}^n a_{i,2} y^*_i.$ By using the same argument as in the proof of the above claim, we can show that there exists integers $\left\{z_i^*\right\}_{i=1}^n$ such that $\left\{y_i^*,z_i^*\right\}_{i=1}^n$ is a solution of problem \eqref{issp}.

{Case B:} there does not exist binary $\left\{\bar y_i\right\}_{i=1}^n$ such that \eqref{lowerupper} holds true. This means that, for any binary $\left\{y_i\right\}_{i=1}^n$ satisfying $\sum_{i=1}^n a_{i,1} y_i \le T$, we must have $\sum_{i=1}^n a_{i,2} y_i < T.$ Therefore, for any binary $\left\{y_i\right\}_{i=1}^n,$ we have
\be\label{equivalenceset}\sum_{i=1}^n a_{i,1} y_i \le T\Longleftrightarrow\sum_{i=1}^n a_{i,2} y_i < T.
\ee
 Hence, the optimal value of problem \eqref{issp_01} is strictly less than $T$ and it is equivalent to
\be \label{issp_tmp2}
 \ba{cl}
 \max\limits_{y} & \displaystyle \sum_{i=1}^n a_{i,2} y_i \\
  \mbox{s.t.} & \displaystyle \sum_{i=1}^n a_{i,1} y_i \le T, \\
       & y_i \in \{0, 1\},~i=1,2,\ldots,n.
 \ea
\ee Moreover, {the relation} \eqref{equivalenceset} implies that the solution of problem \eqref{issp} must satisfy $z_i=y_i\left(a_{i,2}-a_{i,1}\right)$ for all $i=1,2,\ldots,n,$ {since this maximizes the objective without violating the constraints.}   Combining this and \eqref{equivalenceset}, we know that problem \eqref{issp} is equivalent to problem \eqref{issp_tmp2}.

{Combining Cases} {A} and {B}, we can conclude that ISSP \eqref{issp} is equivalent to problem (\ref{issp_01}). This completes the proof.\qed \end{proof}

\subsection*{Proof of Theorem \ref{thm:poly1}}
\begin{proof}
 We prove the theorem by considering the following two cases. 

{Case A:} $T \ge \sum_{i=1}^{n} a_{i,1}.$ In this case, it is simple to find the solution of ISSP (\ref{issp}): if $T\leq \sum_{i=1}^{n} a_{i,2},$ then the solution to ISSP (\ref{issp}) is $y_i=1$ for all $i=1,2,\ldots,n$ and $z_i$ is given by \eqref{zi} with $\bar I$ there being replaced with $n;$ otherwise the solution to ISSP (\ref{issp}) is $y_i=1$ and $z_i=a_{i,2}-a_{i,1}$ for all $i=1,2,\ldots,n.$

{Case B:} $T < \sum_{i=1}^{n} a_{i,1}.$ Then, we can find $I\ge 0$ such that \be \label{ineq1} \sum_{i=1}^{I} a_{i,1} \le T < \sum_{i=1}^{I+1} a_{i,1}\ee in polynomial time. If \be\label{ineq2}\sum_{i=1}^I a_{i,2}>T,\ee then we can easily construct a solution such that the optimal value of ISSP (\ref{issp}) is $T$ in polynomial time as in Case {A} of the proof of Theorem \ref{thm:issp_and_knapsack} and thus ISSP (\ref{issp}) is polynomial time solvable. Next, we show {the truth of \eqref{ineq2} under the assumption (\ref{condition})}.

From the right hand side of \eqref{ineq1}, we get $$T < \sum_{i=1}^{I+1} a_{i,1} \le (I+1) \max\limits_{1 \le i \le n} \{ a_{i,1} \},$$ which further implies
$$I \ge \left\lfloor \dfrac{T}{\max\limits_{1 \le i \le n} \{ a_{i,1} \} } \right\rfloor.$$
Combining {this} with \eqref{condition} yields
\be I \ge  \left\lceil \dfrac{ \max\limits_{1 \le i \le n} \{ a_{i,1} \} }{ \min\limits_{1\le i \le n} \{ a_{i,2} - a_{i,1} \} } \right\rceil, \ee
which means \be \label{duetoassumption} I  \min\limits_{1\le i \le n} \{ a_{i,2} - a_{i,1} \} \ge  \max\limits_{1 \le i \le n} \{ a_{i,1} \}.\ee
Now, we can use \eqref{ineq1} and \eqref{duetoassumption} to obtain \eqref{ineq2}. In particular, we have
\begin{align*}
\sum_{i=1}^{I} a_{i,2}&= \sum_{i=1}^{I} (a_{i,2} - a_{i,1}) + \sum_{i=1}^{I} a_{i,1} \\
                             &\ge I  \min_{1\le i \le n} \{ a_{i,2} - a_{i,1} \} + \sum_{i=1}^{I} a_{i,1} \\
&\ge \max_{1 \le i \le n} \{ a_{i,1} \} + \sum_{i=1}^{I} a_{i,1} \\
&> T,
\end{align*} where the second inequality is due to \eqref{duetoassumption} and the last inequality is due to the right hand side of \eqref{ineq1}. This completes the proof of Theorem \ref{thm:poly1}. \qed \end{proof}

{\subsection*{Proof of Theorem \ref{thm:ktimes-issp}}}
\begin{proof}
  Without loss of generality, assume $a_{1,2} = \max\limits_{1\le i\le n} a_{i,2}$ in this proof. We have the following result.
\begin{lemma}\label{lem:ktimes-issp}
 Suppose \eqref{ratio} holds true for some $c > 1$ and $T$ obeys the uniform distribution over the interval $\left(a_{1,2}, \displaystyle\sum_{i=1}^n a_{i,2}\right].$ Then, the probability that ISSP (\ref{prob:issporigin}) is polynomial time  solvable is at least $$\dfrac{ \sum_{j=2}^n \min \left\{ \sum_{i=1}^{j} \left(1-\dfrac{1}{c}  \right)a_{i,2}, a_{j,2} \right\} }{  \sum_{j=2}^n  a_{j,2} }.$$
\end{lemma}
Combining Lemma \ref{lem:ktimes-issp} and the following inequality
\begin{align*}
\sum\limits_{j=2}^n \min \left\{ \sum\limits_{i=1}^{j} \left(1-\dfrac{1}{c}  \right)a_{i,2}, a_{j,2} \right\}\geq& \sum\limits_{j=2}^n \min \left\{ 2 \left(1-\dfrac{1}{c}  \right)a_{j,2}, a_{j,2} \right\},\\
                                = & \min \left\{ 2 \left(1-\dfrac{1}{c}  \right),1\right\} \sum\limits_{j=2}^n  a_{j,2},
\end{align*} we immediately obtain Theorem \ref{thm:ktimes-issp}.

Next, we show the truth of Lemma \ref{lem:ktimes-issp}. The interval $\left(a_{1,2}, \sum\limits_{i=1}^n a_{i,2}\right]$ can be partitioned as follows:
\be \left(a_{1,2}, \sum\limits_{i=1}^n a_{i,2}\right] = \bigcup\limits_{j=2}^n \left( \sum\limits_{i=1}^{j-1} a_{i,2}, \sum\limits_{i=1}^{j} a_{i,2} \right].
\ee As shown in Case {A} of Theorem \ref{thm:issp_and_knapsack}, for any $T \in \left[\sum\limits_{i=1}^j a_{i,1}, \sum\limits_{i=1}^{j} a_{i,2} \right]$ with $j\in \{1,2,\ldots,n\}$, ISSP (\ref{prob:issporigin}) is polynomial time solvable. Therefore, the probability that ISSP (\ref{prob:issporigin}) is polynomial time solvable is greater than or equal to
\be\label{probability}\dfrac{ \left| \bigcup\limits_{j=2}^n  \left(  \left[\sum\limits_{i=1}^j a_{i,1}, \sum\limits_{i=1}^{j} a_{i,2} \right] \bigcap \left( \sum\limits_{i=1}^{j-1} a_{i,2}, \sum\limits_{i=1}^{j} a_{i,2} \right]  \right) \right| }{  \left| \bigcup\limits_{j=2}^n  \left( \sum\limits_{i=1}^{j-1} a_{i,2}, \sum\limits_{i=1}^{j} a_{i,2} \right] \right|},
\ee where $|\cdot|$ denotes the length of the corresponding {set}. Moreover, it can be verified that the denominator of \eqref{probability} is equal to $\sum\limits_{j=2}^n  a_{j,2}$ and the numerator of \eqref{probability} is lower bounded by $\sum\limits_{j=2}^n \min \left\{ \sum\limits_{i=1}^{j} \left(1-\dfrac{1}{c}  \right)a_{i,2}, a_{j,2} \right\}.$ This completes {the proof} of Lemma \ref{lem:ktimes-issp}.\qed
\end{proof}

\subsection*{Proof of Lemma \ref{lem:neworder}}
\begin{proof}
By Lemma \ref{lem:canonicalsolution}, the ISSP has an optimal solution with at most one midrange element and all left anchored intervals precede the midrange interval and all right anchored intervals follow the midrange interval. Without loss of generality, suppose $\left\{x_i^*\right\}_{i=1}^n$ is such an optimal solution with $[a_{j,1}, a_{j,2}]$ being the only midrange interval (i.e., $x_j^*\in(a_{j,1}, a_{j,2})$) and $[a_{k,1},a_{k,2}]$ with $k>j$ being the last right anchored interval (i.e., $x_k^*=a_{k,2}$). 
Next, we construct an optimal solution $\left\{\tilde x_i^*\right\}_{i=1}^n$ as follows:
\begin{equation*}
\tilde x_{i}^*=\left\{\begin{array}{ll}
x_i^*,&\text{if~}i\neq j~\text{or}~k,\\
a_{k,2}-a_{j,2}+x_j^*, &\text{if~}i=k,\\
a_{j,2}, &\text{if~}i=j.\\
\end{array}
\right.
\end{equation*}
Since $a_{j,2}-a_{j,1}\leq a_{k,2}-a_{k,1}$ and $x_j^*\in(a_{j,1},a_{j,2}),$ it follows that 
$$a_{k,1}=a_{k,2}-(a_{k,2}-a_{k,1}) \leq a_{k,2}-(a_{j,2}-a_{j,1})<a_{k,2}-(a_{j,2}-x_{j}^*)<a_{k,2}.$$ Therefore, $\left\{\tilde x_i^*\right\}_{i=1}^n$ constructed in the above is feasible and optimal to the ISSP. Moreover, the solution $\left\{\tilde x_i^*\right\}_{i=1}^n$ contains only one midrange element $\tilde x_k^*$ and there are neither left nor right anchored intervals following this midrange interval. The proof is completed.\qed\end{proof}

\subsection*{Proof of Lemma \ref{lem:relaxeddp-approx}}
\begin{proof}
We prove the lemma by induction. Obviously, the lemma is true for $i=1.$ Assume it is true for some $i\geq 1.$ Next, we show it is also true for $i+1.$ By the assumption and the fact $\Delta^*_i \subseteq \Delta^*_{i+1},$ we only need to consider the elements $\delta$ in the set 
\be \label{setstar}
\Delta^*_{i+1}\setminus \Delta^*_{i}=\left\{\delta + a_{i+1,1}, \delta + a_{i+1,2} ~|~ \delta \in \Delta_{i}^* \}  \cup \{ a_{i+1,1}, a_{i+1,2} \right\}.
\ee The lemma is trivially true if $\delta=a_{i+1,1}$ or $\delta=a_{i+1,2}.$ It remains to show that the lemma is true for $\delta=\delta'+v\leq \tilde T$ where $\delta'\in\Delta^*_{i}$ and $v\in\{a_{i+1,1}, a_{i+1,2}\}.$ According to the assumption, we divide the subsequent proof into two Cases A and B.

Case A: there exist $\underline{\delta'},~\overline{\delta'} \in \Delta_i$ such that \be\label{assumption1}\underline{\delta'} \le \delta' \le \overline{\delta'}~\text{and}~\overline{\delta'} - \underline{\delta'} \le \epsilon T.\ee In this case, we further consider the following three subcases A1, A2, and A3.

\begin{enumerate}
\item [A1:] $\underline{\delta'}+v \in I_j$ and $\overline{\delta'}+v \in I_j$ for some $j$. Then, let $\underline\delta$ and $\overline\delta$ be the minimum and maximum values of $\Delta_{i+1}$ in the subinterval $I_j,$ respectively. It is simple to check that
    $\underline\delta \le \underline{\delta'}+v \le \delta \le  \overline{\delta'}+v \le \overline\delta$ and $\overline\delta - \underline\delta \le \epsilon T.$

\item [A2:] $\underline{\delta'}+v \in I_j$ and $\overline{\delta'}+v \in I_{j+1}$ for some $j$. Let $\underline{\delta_j}$ and $\overline{\delta_j}$ ($\underline{\delta_{j+1}}$ and $\overline{\delta_{j+1}}$) be the minimum and maximum values of $\Delta_{i+1}$ in the subinterval $I_j$ ($I_{j+1}$), respectively. Then, by \eqref{assumption1} and the assumption in the subcase A2, there must exist $\underline{\delta_j},\,\overline{\delta_j},\,\underline{\delta_{j+1}},\,\overline{\delta_{j+1}} \in \Delta_{i+1}$ such that $\underline{\delta_j} \le \underline{\delta'}+v \le \overline{\delta_j} \le \underline{\delta_{j+1}} \le \overline{\delta'}+v \le \overline{\delta_{j+1}}$, $\overline{\delta_{j}} - \underline{\delta_{j}} \le \epsilon T$, $\underline{\delta_{j+1}} - \overline{\delta_j} \le ( \overline{\delta'}+v)  - ( \underline{\delta'}+v ) \le \epsilon T$, and $\overline{\delta_{j+1}} - \underline{\delta_{j+1}}  \le \epsilon T.$ 
    If $\delta'+v\in [\underline{\delta_j}, \overline{\delta_j}],$ let $\underline\delta=\underline{\delta_j}$ and $\overline\delta=\overline{\delta_j};$ if $\delta'+v\in [\overline{\delta_j}, \underline{\delta_{j+1}}],$ let $\underline\delta=\overline{\delta_j}$ and $\overline\delta=\underline{\delta_{j+1}};$ and if $\delta'+v\in [\underline{\delta_{j+1}}, \overline{\delta_{j+1}}],$ let $\underline\delta=\underline{\delta_{j+1}}$ and $\overline\delta=\overline{\delta_{j+1}}.$ It is simple to verify that $\underline\delta$ and $\overline\delta$ constructed in the above satisfy 
    $\underline\delta \le \delta \le \overline\delta$ and $\overline\delta - \underline\delta \le \epsilon T.$

\item [A3:] $\underline{\delta'}+v \in I_j$ for some $j$ and $\overline{\delta'}+v > \tilde T.$ Since $\tilde T-\epsilon T$ and $\tilde T$ are in different subintervals, we assume $\tilde T-\epsilon T\in I_j$ and $\tilde T\in I_{j+1}$ without loss of generality. Then, we get either $\delta'+v\in I_j$ or $\delta'+v\in I_{j+1}.$ If $\delta'+v\in I_j,$ let $\underline\delta$ and $\overline\delta(\leq \tilde T)$ to be the minimum and maximum values of $\Delta_{i+1}$ in the subinterval $I_j,$ respectively, and we thus have
    $\underline\delta \le \delta \le  \overline\delta$ and $\overline\delta - \underline\delta \le \epsilon T.$ If $\delta'+v\in I_{j+1},$ let $\underline{\delta}$ be the minimum value of $\Delta_{i+1}$ in $I_{j+1}$. Since $\tilde T-\epsilon T\in I_j,$ it follows that $\tilde T - \epsilon T \le \underline\delta \le \delta \le \tilde T.$
\end{enumerate}

 Case B: there exists $\underline{\delta'} \in \Delta_i$ such that $\tilde T - \epsilon T \le \underline{\delta'} \le \delta' \le \tilde T.$ Without loss of generality, assume $\tilde T-\epsilon T\in I_j$ and $\tilde T\in I_{j+1}$ for some $j.$ We can show that the lemma is true for this case by using the same argument as in the above subcase A3. More specifically, we can show that: if $\delta\in I_j,$ there exist $\underline\delta,~\overline\delta \in \Delta_{i+1}$ such that $\underline\delta \le \delta \le  \overline\delta$ and $\overline\delta - \underline\delta \le \epsilon T;$ and if $\delta\in I_{j+1},$ there exists $\underline{\delta}$ such that  $\tilde T - \epsilon T \le \underline\delta \le \delta \le \tilde T.$

This completes the proof of Lemma \ref{lem:relaxeddp-approx}.\qed \end{proof}

\subsection*{Proof of Corollary \ref{cor:relaxeddp-approx}}
\begin{proof}
For $\delta^* \in \Delta_{\tilde \Lambda}^*,$ by Lemma \ref{lem:relaxeddp-approx}, we have one of the following two statements:
\begin{enumerate}
  \item there exist $\underline\delta,\,\overline\delta \in \Delta_{\tilde \Lambda}$ such that $\underline\delta \le \delta^* \le \overline\delta$ and $\overline\delta - \underline\delta \le \epsilon T;$ 
  \item there exists $\underline\delta \in \Delta_{\tilde \Lambda}$ such that $\tilde T - \epsilon T \le \underline\delta \le \delta^* \le \tilde T$.
\end{enumerate}
If the first statement is true, let $\delta$ be $\overline\delta$ there. Since $\delta^*$ is the optimal value of the ISSP, we must have $\delta=\delta^*.$ If the second statement is true, let $\delta$ be $\underline\delta$ there. Then, we immediately obtain the desired result.\qed \end{proof}

\subsection*{Proof of Lemma \ref{lem:divideinto2parts}}
\begin{proof}
 Since $\delta \in \Delta_{\tilde \Lambda}\subseteq \Delta_{\tilde \Lambda^*}$, it follows that there must exist $\delta_1 \in \left\{0\right\}\cup\Delta_{\tilde \Lambda_1}^*$ and $\delta_2 \in \left\{0\right\}\cup\Delta_{\tilde \Lambda_2}^*$ such that $\delta_1 + \delta_2 = \delta$. Invoking Lemma \ref{lem:relaxeddp-approx} again, we have one of the following two statements: \begin{enumerate}
    \item there exists $\underline{\delta_1} \in \Delta_{\tilde \Lambda_1}$ such that $\tilde T - \epsilon T \le \underline{\delta_1}\leq \delta_1\leq \tilde T$ or there exists $\underline{\delta_2} \in \Delta_{\tilde \Lambda_2}$ such that $\tilde T - \epsilon T \le \underline{\delta_2}\leq \delta_2\leq \tilde T;$
    \item there exist $\underline{\delta_{1}} \in \Delta_{\tilde \Lambda_1}, \overline{\delta_1}  \in \Delta_{\tilde \Lambda_1} , \underline{\delta_2} \in \Delta_{\tilde \Lambda_2}, \overline{\delta_2}  \in \Delta_{\tilde \Lambda_2}$ such that $\underline{\delta_{1}} \le \delta_1 \le \overline\delta_{1},$ $\underline{\delta_{2}} \le \delta_2 \le \overline{\delta_{2}},$ $0\le \overline{\delta_{1}} - \underline{\delta_{1}} \le \epsilon T$, and $0\le \overline{\delta_{2}} - \underline{\delta_{2}} \le \epsilon T.$
  \end{enumerate} If the first statement is true, then let $(u_1,u_2)=(\underline{\delta_1},0)$ or $(u_1,u_2)=(0,\underline{\delta_2})$. Obviously, $u_1$ and $u_2$ defined in the above satisfy $\tilde T - \epsilon T \le u_1 + u_2 \le \tilde T.$ It remains to show the lemma if the second statement is true. We consider the following three cases separately.

 Case {A}: $\underline{\delta_{1}}+\underline{\delta_{2}} \geq \tilde T - \epsilon T.$ In this case, let $(u_1,u_2)=(\underline{\delta_{1}},\underline{\delta_2}).$ Then, combining the facts $\delta_1 + \delta_2 = \delta,$ $\underline{\delta_{1}} \le \delta_1,$ $\underline{\delta_{2}} \le \delta_2$ and $\delta \le \tilde T,$ we immediately obtain $\tilde T - \epsilon T\leq u_1+u_2=\underline{\delta_{1}}+\underline{\delta_{2}}\leq \delta_1+\delta_2=\delta\leq \tilde T.$

  Case {B}: $\overline{\delta_{1}}+\overline{\delta_{2}} \leq \tilde T.$ In this case, let $(u_1,u_2)=(\overline{\delta_{1}},\overline{\delta_2}).$ We can use essentially the same argument as in the above Case {A} to show $\tilde T - \epsilon T\leq u_1+u_2\leq \tilde T.$

  Case {C}: $\underline{\delta_{1}}+\underline{\delta_{2}} < \tilde T - \epsilon T$ and $\overline{\delta_{1}}+\overline{\delta_{2}} > \tilde T.$ In this case, let $(u_1,u_2) = (\overline{\delta_{1}},\underline{\delta_{2}}).$ Combining the conditions assumed in this case, $\overline{\delta_{1}} - \underline{\delta_{1}} \le \epsilon T,$ and $\overline{\delta_{2}} - \underline{\delta_{2}} \le \epsilon T,$ we immediately obtain $u_1+u_2=\overline{\delta_{1}} + \underline{\delta_{2}} = (\overline{\delta_{1}} + \overline{\delta_{2}}) - (\overline{\delta_{2}} -\underline{\delta_{2}} ) > \tilde T - \epsilon T$ and $u_1+u_2=\overline{\delta_{1} } + \underline{\delta_{2}} = (\underline{\delta_{1}} + \underline{\delta_{2}}) + (\overline{\delta_{1}} -\underline{\delta_{1}} ) < \tilde T$.

 From the above analysis, we conclude that there exist $u_1 \in \{0\} \cup \Delta_{\tilde \Lambda_1}$ and $u_2 \in \{0\} \cup \Delta_{\tilde \Lambda_2}$ such that $\tilde T - \epsilon T \le u_1 + u_2 \le \tilde T$. The proof is completed.\qed\end{proof}

\subsection*{Proof of Lemma \ref{lem:backtracking}}
\begin{proof} By the assumption, the largest number in $\Delta^*_{\tilde \Lambda}$ associated with the target $\tilde T$ must be in the interval $[\tilde T - \epsilon T, \tilde T]$. Then, it follows from Corollary \ref{cor:relaxeddp-approx} that the largest number in $\Delta_{\tilde \Lambda}$ associated with the target $\tilde T$ must also lie in the interval $[\tilde T - \epsilon T , \tilde T].$ Recall the procedure \textbf{backtracking}, we know the following facts: when the procedure starts, the largest number in $\Delta_{\tilde \Lambda}$ associated with the target $\tilde T$ (i.e., $u=\max\{\delta^+(j), \delta^-(j) ~|~  \delta^+(j)\le \tilde T, \delta^-(j) \le \tilde T \}$) is found in line \ref{line:init_u}; the recent intervals which contribute to generate $u$ are backtracked in lines \ref{line:begin_main_backtrack} -- \ref{line:end_speedup_backtrack}; and when the procedure terminates, $y+u$ is in the interval $[\tilde T - \epsilon T, \tilde T].$ Since $u$ is generated by the procedure \textbf{relaxed dynamic programming}, it follows that $u \in \{ 0 \} \cup \Delta_{\tilde \Lambda \setminus \Lambda^E}.$ This further implies that there exists $\delta \in \{ 0 \} \cup \Delta^*_{\tilde \Lambda \setminus \Lambda^E}$ such that $\tilde T - \epsilon T \le y + \delta \le \tilde T.$ This completes the proof of Lemma \ref{lem:backtracking}.\qed\end{proof}

\subsection*{Proof of Lemma \ref{dciscorrect}}
\begin{proof} First, as argued in the proof of Lemma \ref{lem:backtracking}, we know that the largest number in $\Delta_{\tilde \Lambda}$ associated with the target $\tilde T$ must also lie in the interval $[\tilde T - \epsilon T , \tilde T].$ By Lemma \ref{lem:divideinto2parts}, we can split $\tilde \Lambda$ into $\tilde \Lambda_1$ and $\tilde \Lambda_2$ as in lines \ref{line:conquer1} and \ref{line:conquer2} of the procedure \textbf{divide and conquer}, and find $u_1 \in \{0\} \cup \Delta_{\tilde \Lambda_1}$ and $u_2 \in \{0\} \cup \Delta_{\tilde \Lambda_2}$ such that $\tilde T - \epsilon T \le u_1 + u_2 \le \tilde T$. Without loss of generality, assume both $u_1$ and $u_2$ are positive. Otherwise, if $u_1=0$~($u_2=0$), then we can remove the intervals in $\tilde \Lambda_1$ ($\tilde \Lambda_2$) and split $\tilde \Lambda_2$ ($\tilde \Lambda_1$) again. This implies that there exists positive $u_1 \in \Delta^*_{\tilde \Lambda_1}$ satisfying $u_1 \in [\tilde T - u_2 - \epsilon T, \tilde T - u_2].$ Moreover, by Lemma \ref{lem:backtracking}, we know that there exists $\delta \in \{0\} \cup \Delta^*_{\tilde \Lambda_1 \setminus \Lambda^E}$ such that $\delta \in [\tilde T-u_2-y_1^B-\epsilon T, \tilde T -u_2-y_1^B],$ where $y_1^B\geq 1$ is the output of line \ref{line:dc_backtrack1} of the procedure \textbf{divide and conquer}. This in turn shows that the assumption of Lemma \ref{dciscorrect} is satisfied for the procedure \textbf{divide and conquer} in line \ref{line:dc_dc1}. Since at least one interval is removed after each recursive call, the recursive calls of the procedure \textbf{divide and conquer} will eventually end. Consequently, we get $y_1^{DC} \in [\tilde T - u_2 - y_1^B - \epsilon T, \tilde T - u_2 - y_1^B]$, which is equivalent to $u_2 \in [\tilde T - y_1^B - y_1^{DC} - \epsilon T, \tilde T - y_1^B - y_1^{DC} ]$. Similar analysis applies to lines \ref{line:conquer3}, \ref{line:dc_backtrack2}, and \ref{line:dc_dc2} of the procedure \textbf{divide and conquer}. This completes the proof of Lemma \ref{dciscorrect}.\qed \end{proof}

\subsection*{Proof of Theorem \ref{thm1}}
\begin{proof}
By enumerating all intervals $\left\{[a_{i,1},a_{i,2}]\right\}_{i=1}^n,$ Algorithm \ref{algo:issp-fptas} can successfully find the (possible) midrange interval $[a_{m,1},a_{m,2}]$ and the largest number $\hat \delta$ in $\Delta_{\Lambda\setminus\Lambda^E}$ associated with the target $T-{a_{m,1}}.$ Suppose that $\delta^*$ is the largest number in $\Delta^*_{\Lambda\setminus\Lambda^E}$ associated with the target $T-{a_{m,1}}.$ Then, by Lemma \ref{lem:relaxeddp-approx}, we have either $\hat \delta=\delta^*$ or $T-{a_{m,1}}-\epsilon T\leq \hat \delta\leq \delta^*\leq T-{a_{m,1}}.$ We consider the following four cases.

Case {A}: $\hat \delta+\epsilon T<T-{a_{m,1}}$ and $\hat \delta=\delta^*.$ In this case, we have $\hat \delta \leq \delta^*\leq \hat \delta+\epsilon T.$ Using Lemma \ref{dciscorrect}, we immediately obtain $\hat \delta\leq \hat T^A\leq \hat \delta+\epsilon T,$ where $\hat T^A$ is the output the procedure \textbf{divide and conquer} $\left(\Lambda \setminus \Lambda^E, \hat \delta+\epsilon T\right)$ in line \ref{line:call_dc} of Algorithm \ref{algo:issp-fptas}. Since $\hat\delta=\delta^*$ is the largest number in $\Delta^*_{\Lambda\setminus\Lambda^E}$ associated with the target $T-{a_{m,1}},$ it follows that $\hat T^A=\hat\delta=\delta^*.$ Hence, $$T^A=\hat T^A+\min\left\{a_{m,2},T-\hat T^A\right\}=\delta^*+\min\left\{a_{m,2},T-\delta^*\right\}$$ is the optimal value of the ISSP.

Case {B}: $\hat \delta+\epsilon T<T-{a_{m,1}}$ and $T-{a_{m,1}}-\epsilon T\leq \hat \delta\leq \delta^*\leq T-{a_{m,1}}.$ This case will not happen, since the two conditions contradict with each other.

Case {C}: $\hat \delta+\epsilon T\geq T-{a_{m,1}}$ and $\hat \delta=\delta^*.$ From these two conditions and the fact that $\delta^*$ is the largest number in $\Delta^*_{\Lambda\setminus\Lambda^E}$ associated with the target $T-{a_{m,1}},$ we obtain $\delta^*\in[T-{a_{m,1}}-\epsilon T, T-{a_{m,1}}].$ Then, by Lemma \ref{dciscorrect}, we know that the returned approximate value of $\hat T^A$ satisfies $T-{a_{m,1}}\geq \hat T^A \geq T-{a_{m,1}}-\epsilon T,$ and $T^A=\hat T^A+\min\left\{a_{m,2},T-\hat T^A\right\}\geq \hat T^A+a_{m,1}\geq T-\epsilon T.$

Case {D}: $\hat \delta+\epsilon T\geq T-{a_{m,1}}$ and $T-{a_{m,1}}-\epsilon T\leq \hat \delta\leq \delta^*\leq T-{a_{m,1}}.$ The same argument as in the above Case {C} shows that $T-{a_{m,1}}\geq \hat T^A \geq T-{a_{m,1}}-\epsilon T$ and $T^A\geq T-\epsilon T.$

From the above analysis, we can conclude that Algorithm \ref{algo:issp-fptas} either returns an optimal solution, or an approximate solution with the objective value being great than or equal to $T-\epsilon T$. The proof is completed. \qed \end{proof}

\subsection*{Proof of Theorem \ref{thm2}}
\begin{proof}We analyze the time and space complexities of Algorithm \ref{algo:issp-fptas} separately. We first consider the time complexity of Algorithm \ref{algo:issp-fptas}. The time complexity of sorting $n$ intervals by length (line \ref{sort} of Algorithm \ref{algo:issp-fptas}) is ${\cal O}(n \log n)$.
The procedure \textbf{relaxed dynamic programming} is called many times in Algorithm \ref{algo:issp-fptas} and performing the procedure \textbf{relaxed dynamic programming} is the dominated computational cost in the recursive framework of the procedure \textbf{divide and conquer}. It is simple to see that the time complexity of performing the procedure \textbf{relaxed dynamic programming} with inputs $(\tilde \Lambda,\tilde T)$ is ${\cal O}\left(\tilde n \tilde l\right),$ where $\tilde n=|\tilde \Lambda|$ and $\tilde l=\left\lceil \dfrac{ \tilde T }{ \epsilon T } \right\rceil.$

Now, we bound the times that the procedure \textbf{divide and conquer} is performed. To do so, let us denote the root node of the recursive tree as level $0.$ Then, there are at most $2^l\leq n$ nodes in the $l$-th level of the recursive tree. For ease of presentation, we assume that there are $2^l$ nodes in the $l$-th level of the recursive tree and denote the targets of these $2^l$ nodes by $\tilde T_{l,1},$ $\tilde T_{l,2},$ $\ldots, \tilde T_{l,2^l}$ and item sets of these $2^l$ nodes by $\tilde \Lambda_{l,1},$ $\tilde \Lambda_{l,2},$ $\ldots, \tilde \Lambda_{l,2^l}$ for all $l=1,2,\ldots,\lceil\log n\rceil.$ Then, we must have $\sum\limits_{i=1}^{2^l} \tilde T_{l,i} \le T$ for all $l=1,2,\ldots,\lceil\log n\rceil$ and $\left|\tilde \Lambda_{l,i}\right|={\cal O}\left(\frac{n}{2^l}\right)$ for all $i=1,2,\ldots,2^l$ and $l=1,2,\dots, \lceil\log n\rceil.$ Therefore, the total time complexity of calling the procedure \textbf{divide and conquer} in Algorithm \ref{algo:issp-fptas} is
$$\sum_{l=0}^{\lceil\log n\rceil}\sum_{i=1}^{2^l}{\cal O}\left(\left\lceil\frac{\tilde T_{l,i}}{\epsilon T}\right\rceil\left|\Lambda_{l,i}\right|\right)\leq {\cal O}(n/\epsilon),$$
and the total time complexity of  Algorithm \ref{algo:issp-fptas} is ${\cal O}\left(n \max\left\{1 / \epsilon,\log n\right\}\right).$

Next, we consider the space complexity of Algorithm \ref{algo:issp-fptas}. It takes ${\cal O}(n)$ space complexity to store the interval set $\left\{[a_{i,1}, a_{i,2}]\right\}_{i=1}^n.$ The space complexity required to store the relaxed dynamic programming arrays $\delta_1^-(\cdot)$, $\delta_1^+(\cdot)$, $\delta_2^-(\cdot)$, $\delta_2^+(\cdot)$, $d_{1,1}(\cdot)$, $d_{1,2}(\cdot)$ $d_{2,1}(\cdot)$, $d_{2,2}(\cdot)$ is ${\cal O}(1/\epsilon)$. Since the memory space can be reused in the recursive calls of the procedure \textbf{divide and conquer}, we conclude that the total space complexity of Algorithm \ref{algo:issp-fptas} is ${\cal O}(n+1/\epsilon).$ \qed\end{proof}

{\section*{Appendix B: An Illustration of Algorithm \ref{algo:issp_pseudopoly} and Algorithm \ref{algo:issp-fptas}}}
To make Algorithm \ref{algo:issp_pseudopoly} and Algorithm \ref{algo:issp-fptas} clear, an illustration of applying them to solve the following ISSP instance is given: $$T=100,~n=4,$$$$[a_{1,1},a_{1,2}]=[10,20],~[a_{2,1},a_{2,2}]=[10,25],$$$$[a_{3,1},a_{3,2}]=[60,85],~[a_{4,1},a_{4,2}]=[20,50].$$

If Algorithm \ref{algo:issp_pseudopoly} is applied to solve the above instance:
when $i=1,$ executing lines \ref{line:forstart}--\ref{line:dynamic} gives $$\delta^*=0,~T^*=20,~m=1,~\Delta_1^*=\left\{10,20\right\};$$ then $i=2$ and executing lines \ref{line:forstart}--\ref{line:dynamic} gives $$\delta^*=20,~T^*=45,~m=2,~\Delta_2^*=\left\{10,20,25,30,35,45\right\};$$ then $i=3$ and executing lines \ref{line:forstart}--\ref{line:dynamic} gives $\delta^*=35,$ $T^*=100,$ and $~m=3.$
Since $T^*=T=100$ when $i=3,$  Algorithm \ref{algo:issp_pseudopoly} goes to line \ref{line:deltastar} directly without computing 
$$\Delta_3^*=\left\{10,20,25,30,35,45,60,70,80,85,90,95\right\}.$$
Then, executing lines \ref{line:deltastar}--\ref{line:solution3} gives $\delta^*=35$ and returns the optimal solution
$$x_1^*=10,~x_2^*=25,~x_3^*=65,~x_4^*=0.$$ It can be seen that $[a_{1,1},a_{1,2}]$ is a left anchored interval, $[a_{2,1},a_{2,2}]$ is a right anchored interval, $[a_{3,1},a_{3,2}]$ is the only midrange interval, and there is no left/right anchored intervals following the midrange interval. Therefore, the returned solution by Algorithm \ref{algo:issp_pseudopoly} satisfies the property in Lemma \ref{lem:neworder}.

If Algorithm \ref{algo:issp-fptas} with $\epsilon=0.2$ (and thus $l=5$) is applied to solve the above instance:
when $i=1,$ executing lines \ref{line:beginfor}--\ref{line:endfor} gives $$\bar\delta=0,~\hat T=20,~\hat \delta=0,~m=1,~\tilde \Delta=\left\{10,20\right\},~\delta^{-}(1)=10,~\delta^{+}(1)=20,$$ and $$\delta^{-}(2)=\delta^{+}(2)=\delta^{-}(3)=\delta^{+}(3)=\delta^{-}(4)=\delta^{+}(4)=\delta^{-}(5)=\delta^{+}(5)=0;$$ then $i=2$ and executing lines \ref{line:beginfor}--\ref{line:endfor} gives $$\bar\delta=20,~\hat T=45,~\hat \delta=20,~m=2,~\tilde \Delta=\left\{10,20,25,30,35,45\right\},~\delta^{-}(1)=10,~\delta^{+}(1)=20,$$ and $$\delta^{-}(2)=25,~\delta^{+}(2)=35,~\delta^{-}(3)=\delta^{+}(3)=45,~\delta^{-}(4)=\delta^{+}(4)=\delta^{-}(5)=\delta^{+}(5)=0;$$ then $i=3$ and executing lines \ref{line:beginfor}--\ref{line:endfor} gives $$\bar\delta=35,~\hat T=100,~\hat \delta=35,~m=3.$$
Since $T^*=\hat T=100$ when $i=3,$ Algorithm \ref{algo:issp-fptas} goes to line \ref{line:lambdaE} directly without computing
$$\tilde\Delta=\left\{60,70,80,85,95\right\},~\delta^{-}(1)=10,~\delta^{+}(1)=20,~\delta^{-}(2)=25,~\delta^{+}(2)=35,$$
$$\delta^{-}(3)=45,~\delta^{+}(3)=60,~\delta^{-}(4)=70,~\delta^{+}(4)=80,~\delta^{-}(5)=85,~\delta^{+}(5)=95.$$
Then, executing line \ref{line:lambdaE} gives $\Lambda^{E}=\left\{[60,85],~[20,50]\right\}$ and Algorithm \ref{algo:issp-fptas} calls the procedure \textbf{divide and conquer} $\left(\Lambda\setminus\Lambda^E,~\min\left\{\hat \delta + \epsilon T,T-a_{m,1}\right\}\right),$ where $$\Lambda\setminus\Lambda^E=\left\{[10,20],~[10,25]\right\}$$ and $$\min\left\{\hat \delta + \epsilon T,T-a_{m,1}\right\}=\min\left\{35+0.2*100, 100-60\right\}=40.$$

With these inputs, line \ref{line:dc1} of the procedure \textbf{divide and conquer} gives $\tilde \Lambda_1=\left\{[10,20]\right\}$ and $\tilde \Lambda_2=\left\{[10,25]\right\};$ line \ref{line:conquer1} of the procedure \textbf{divide and conquer} gives
$$\delta_1^{-}(1)=10,~\delta_1^{+}(1)=20,~\delta_1^{-}(2)=\delta_1^{+}(2)=0,$$$$d_{1,1}(\delta_1^{-}(1))=d_{1,2}(\delta_1^{-}(1))=d_{1,1}(\delta_1^{+}(1))=1,~d_{1,2}(\delta_1^{+}(1))=2;$$
line \ref{line:conquer2} of the procedure \textbf{divide and conquer} gives
$$\delta_2^{-}(1)=\delta_2^{+}(1)=10,~\delta_2^{-}(2)=\delta_2^{+}(2)=25,$$
$$d_{2,1}(\delta_2^{-}(1))=d_{2,1}(\delta_2^{+}(1))=2,~d_{2,2}(\delta_2^{-}(1))=d_{2,2}(\delta_2^{+}(1))=1,$$
$$d_{2,1}(\delta_2^{-}(2))=d_{2,2}(\delta_2^{-}(2))=d_{2,1}(\delta_2^{+}(2))=d_{2,2}(\delta_2^{+}(2))=2;$$ line \ref{line:conqueru} of the procedure \textbf{divide and conquer} finds $(u_1,u_2)=(0,25)$ satisfying $u_1\in\left\{0,10,20\right\},$ $u_2\in\left\{0,10,25\right\},$ and $20=\tilde T-\epsilon T\leq u_1+u_2\leq \tilde T=40.$ Since both $\tilde\Lambda_1$ and $\tilde \Lambda_2$ contain only one interval, the procedure \textbf{backtracking} can successfully return the approximate solution. More specifically,
the procedure \textbf{divide and conquer} skips lines \ref{line:dc_backtrack1} and \ref{line:dc_dc1}; executes line \ref{line:dc_backtrack2} and gives $y_2^B=25$ and $\Lambda^E=\left\{[a_{2,1},a_{2,2}]\right\};$ and skips line \ref{line:dc_dc2}.
Then, line \ref{line:call_dc} of Algorithm \ref{algo:issp-fptas} returns $x_1^A=0$,~$x_2^A=25$, and $\hat T^A=25;$ line \ref{line:m} of Algorithm \ref{algo:issp-fptas} returns $x_3^A=75;$ line \ref{line:m+1} of Algorithm \ref{algo:issp-fptas} returns $x_4^A=0;$ and line \ref{line:last} of Algorithm \ref{algo:issp-fptas} returns $T^A=100.$ Again, the returned $(1-\epsilon)$-optimal solution (actually the optimal solution) by Algorithm \ref{algo:issp-fptas} satisfies the property in Lemma \ref{lem:neworder}, i.e, $[a_{2,1},a_{2,2}]$ is a right anchored interval, $[a_{3,1},a_{3,2}]$ is the only midrange interval, and there is no left/right anchored intervals following the midrange interval.

Two remarks on Algorithm \ref{algo:issp_pseudopoly} and Algorithm \ref{algo:issp-fptas} are in order.

First, although Algorithm \ref{algo:issp-fptas} finds an optimal solution for the above ISSP instance, it cannot be guaranteed to do so for the general ISSP. The reason is that Algorithm \ref{algo:issp-fptas} partitions the interval $(0,T]$ into $\lceil1/\epsilon\rceil$ subintervals and stores only the smallest and largest values lying in the subintervals at each iteration. This is sharply different from Algorithm \ref{algo:issp_pseudopoly}, where all values in $\Delta_i^*$ are stored. For the above instance, when $i=2$, we have $$\left\{\delta^{-}(k),\delta^{+}(k)\right\}_{k=1}^{5}=\left\{10,20,25,35,45\right\}\subset\Delta_2^*;$$
and when $i=3,$ we have
$$\left\{\delta^{-}(k),\delta^{+}(k)\right\}_{k=1}^{5}=\left\{10,20,25,35,45,60,70,80,85,95\right\}\subset\Delta_3^*.$$%

Second, as mentioned below Lemma \ref{lem:divideinto2parts}, the pair $(u_1,u_2)$ satisfying the inequality $\tilde T-\epsilon T\leq u_1+u_2\leq \tilde T$ is generally not unique and different choices of the pair $(u_1,u_2)$ might lead to different approximate solutions. For example, $(u_1,u_2)$ in the above instance can also be $(10,10),$ which results in the approximate solution $$x_1^A=10,~x_2^A=10,~x_3^A=80,~x_4^A=0;$$ or can also be $(10,25),$ which results in the approximate solution $$x_1^A=10,~x_2^A=25,~x_3^A=65,~x_4^A=0;$$ or can also be $(20,0),$ which results in the approximate solution $$x_1^A=20,~x_2^A=0,~x_3^A=80,~x_4^A=0.$$

\begin{acknowledgements}
{We thank Prof. Nelson Maculan and Prof. Sergiy Butenko for the useful discussions on this paper.}
\end{acknowledgements}



\bibliography{ISSP}

\bibliographystyle{spmpsci}

\end{document}